%% file: renren.tex
\documentclass{sig-alternate-10pt}
\usepackage{times}
\usepackage{balance}
\usepackage{url}
\usepackage{epsfig}
\usepackage{graphicx}
\usepackage{subfigure}
\usepackage{amsmath}
\usepackage{amssymb} 
\usepackage{amsfonts}
\usepackage{color}
\usepackage{multirow}
\usepackage{algorithm}
\usepackage[utf8]{inputenc}
\usepackage[T1]{fontenc}

\newcommand{\para}[1]{{\vspace{4pt} \bf \noindent #1 \hspace{10pt}}}
\newcommand{\parab}[1]{{\vspace{4pt} \em \noindent #1 \hspace{3pt}}}
\newcommand{\fixme}[1]{{\color{red} #1}}

\newenvironment{packed_itemize}{
\begin{itemize}
 \setlength{\itemsep}{2pt}
 \setlength{\parskip}{0pt}
 \setlength{\parsep}{0pt}
 \setlength{\headsep}{0pt}
 \setlength{\topskip}{0pt}
 \setlength{\topmargin}{0pt}
 \setlength{\topsep}{0pt}
 \setlength{\partopsep}{0pt}
}{\end{itemize}}

\begin{document}

\title{Multi-scale Dynamics in a Massive Online Social Network}
\author{Xiaohan Zhao$^\dag$, Alessandra Sala$^*$, Christo Wilson$^\dag$, 
  Xiao Wang$^\ddag$, Sabrina Gaito$^\S$,\\ Haitao Zheng$^\dag$, Ben Y. Zhao$^\dag$\\
  $^\dag${\tt Department of Computer Science, UC Santa Barbara}\\
  $^*${\tt Bell Labs, Ireland} \hspace{0.5in} $^\ddag${\tt Renren Inc., Beijing, China}\\
  $^\S${\tt Department of Computer Science, Università degli Studi di Milano}}

\newtheorem{theorem}{Theorem}
\newtheorem{definition}{Definition}
\newtheorem{lemma}{Lemma}
\newtheorem{cor}{Corollary}
\newtheorem{fact}{Fact}
\newtheorem{property}{Property}
\newtheorem{remark}{Remark}
\newtheorem{claim}{Claim}

\maketitle

\begin{abstract}
  Data confidentiality policies at major social network providers have
  severely limited researchers' access to large-scale datasets.  The biggest
  impact has been on the study of network dynamics, where researchers have
  studied citation graphs and content-sharing networks, but few have analyzed
  detailed dynamics in the massive social networks that dominate the web
  today.  In this paper, we present results of analyzing detailed dynamics in
  the Renren social network, covering a period of 2 years when the network
  grew from 1 user to 19 million users and 199 million edges. Rather than
  validate a single model of network dynamics, we analyze dynamics at
  different granularities (user-, community- and network-wide) to determine
  how much, if any, users are influenced by dynamics processes at different
  scales.  We observe independent predictable processes at each level, and
  find that while the growth of communities has moderate and sustained impact
  on users, significant events such as network merge events have a strong but
  short-lived impact that is quickly dominated by the continuous arrival of
  new users.
\end{abstract}

\input{intro}

\input{network}

\input{preferential}

\input{community}
\input{netmerge}
\input{related}

\input{conclusion}
\balance

\begin{small}
\bibliographystyle{acm}
\bibliography{zhao,social,han}
\end{small}

\end{document}

%% file: intro.tex
\section{Introduction}
\label{sec:intro}

A number of interrelated processes drive dynamics in social networks.
A deeper understanding of these processes can allow us to better model and predict
structure and dynamics in social networks. In turn, improved models and predictors
have numerous practical implications on the design of
infrastructure, applications, and security mechanisms for social networks.

Details of these dynamic processes are best studied in the context of today's
massive online social networks (OSNs), {\em e.g.} Facebook~\cite{interaction},
LinkedIn~\cite{linkedin}, and Renren~\cite{renren-imc10}.  Unfortunately, the
providers of these networks generally consider their dynamic network data to
be trade secrets, and have few incentives to make such data available for
research.  Instead, studies have analyzed citation networks~\cite{leskovec2005graphs},
content sharing networks~\cite{kumar2005bursty}, and high level statistics of social
networks~\cite{ahn2007analysis}.
Others~\cite{leskovec2008microscopic,mislove2008growth,garg2009evolution}
sought to verify the validity of generative models such as
preferential attachment (PA)~\cite{barabasi1999emergence}.


Our goal is to better understand in detail the evolutionary dynamics in a
social network. This includes not only the initial growth process during a
social network's formation, but also the ongoing dynamics
afterwards, as the network matures.  Much of the prior work in this area,
including generative graph models and efforts to validate
them~\cite{barabasi1999emergence,leskovec2008microscopic,mislove2008growth,garg2009evolution},
has focused on capturing network dynamics as a single process.  In contrast, we
are interested in the question ``how are individual user dynamics influenced by
processes at different scales?''  How much are the dynamics of users influenced by
external forces and events, such as the activities of friends in communities
they belong to, or by large-scale events that occur at the network level?  

In this work, we explore these questions empirically through a detailed
analysis of social network dynamics at multiple scales: at the individual
user level, at the level of user communities, and at the global
network level.  We study a dynamic graph, {\em i.e.} a sequence of detailed timestamped
events that capture the ongoing growth of the Renren online social
network~\cite{renren-imc10}.  With over 220 million users, Renren is the
largest social network in China, and provides functionality similar to
Facebook.  We focus our analysis on the first two years of Renren's growth, from
its first user in November 2005, to December 2007 when it had over 19 million
members. This captures the network's initial burst of growth, as well as a
period of more sustained growth and evolution.  Our anonymized data includes
timestamps of all events, including the creation of 19 million user accounts and 199
million edges.  This dataset is notable because of three features: its scale,
the absolute time associated with each event, and a rare network {\em merge}
event, when the Renren social network merged with its competitor 5Q.com in
December 2006, effectively doubling its size from 600K users to 1.3 million
users in a single day.


Our analysis of network dynamics in the Renren dataset focuses on three
different levels of granularity: nodes, communities, and networks.  At each
level, we search for evidence of impact on user behavior.  Along the way, we
also make a number of intriguing observations about dynamic processes in
network communities and network-wide events.

{\em Individual Nodes.} The creation of links between
individual users has been studied in a number of contexts, and is long
believed to be driven by generative models based on the principle of
preferential attachment, {\em i.e.} users prefer to connect to nodes with
higher degree~\cite{barabasi1999emergence}.  Our goal is to extend the
analysis of this model with respect to two new dimensions.  First,
preferential attachment defines how a sequence of edges are created in
logical order, but how do node dynamics correlate with absolute time?
Second, does the strength of the preferential attachment model strengthen or
weaken as the network grows in scale and matures?

{\em Communities.} Intuitively, the behavior of a user is likely to be
significantly impacted by the actions of her friends in the network. This has
been previously observed in offline social
networks~\cite{zachary1977information}.  Our goal is to empirically determine
if user activity at the level of communities has a real impact on individual
users.  To do so, we first implement a way to define and track the evolution
of user communities over time.  We track the emergence and dissolution of
communities over time, and quantify the correlation of user behavior to the
lifetime, size, and activity level of the communities they belong to.

{\em Networks.}  Finally, we wish to quantify the impact, if any, of
network-level events on individual user behavior.  By network-level events,
we refer to unusual events that affect the entire network, such as the
merging of two distinct social networks recorded in our dataset.  We analyze
user data before and after the merge of the Renren and 5Q social networks,
and quantify the impact of different factors on user behavior, including
duplicate accounts, and user's edge creation preferences over time.

\para{Key Findings.} Our analysis produces several significant findings.
First, we find that nodes (users) are most active in building links
(friendships) shortly after joining the network.  As the network matures,
however, we find that new edge creation is increasingly dominated by existing
nodes in the system, even though new node arrivals is keeping pace with
network growth.  Second, we find that
influence of the preferential attachment model weakens over time, perhaps
reflecting the reduced visibility of each node over time.  As the network
grows in size, users are less likely to be aware of high degree nodes in the
network, and more likely to obey the preferential model with users within a
limited neighborhood.  Third, at the level of user communities, we find that
users in large communities are more active in creating friends. Active nodes
with high degrees tend to join and help form large communities, and their
activity introduces new friends to their neighbors, further encouraging edge
formation within the community.  In addition, we found that a
combination of community structural features can predict the short-term
``death'' of a community with more than 75\% accuracy.  

Finally, in our analysis of the network merge event, we use user activity to
identify duplicate accounts across the networks. Aside from duplicate
accounts, we find that the network merge event has a distinct short-term
impact on user activity patterns.  Users generate a high burst in edge
creation, but the cross-network activity fades and quickly becomes dominated
by edge creation generated by new users.  Overall, this quickly reduces
average distance between the two networks and melds them into a single
indistinguishable network.

%% file: network.tex
\begin{figure*}[t]
  \centering \mbox{ \subfigure[Absolute Network Growth]
    {\includegraphics[width=0.33\textwidth]{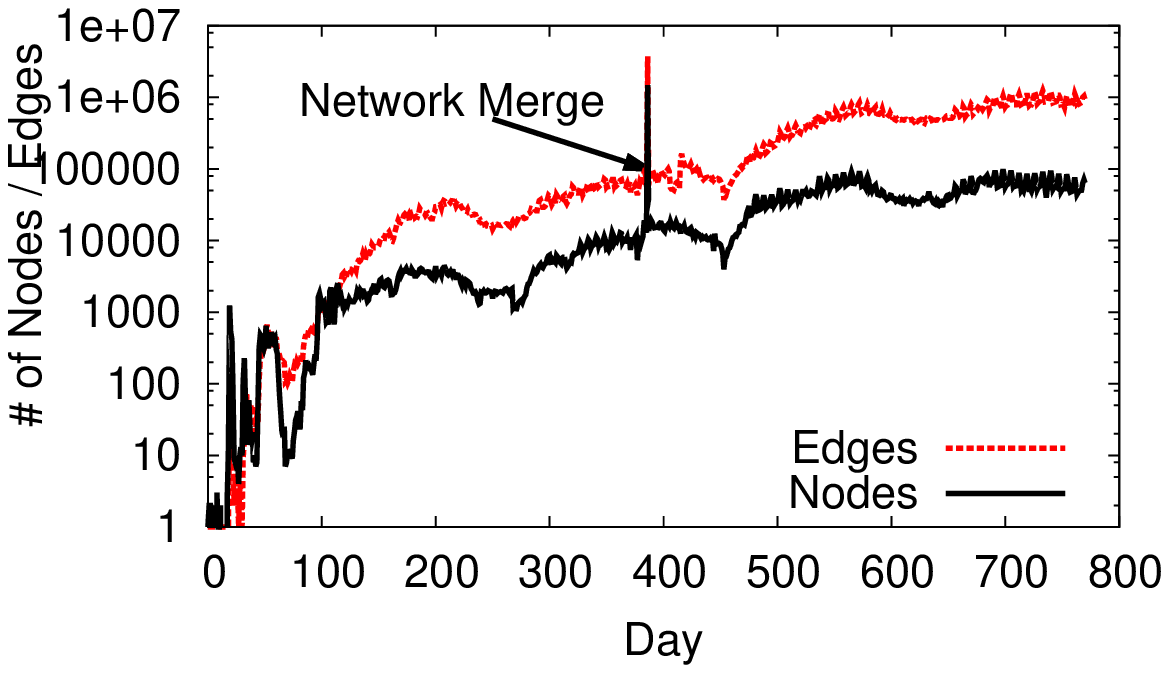}\label{fig:new}}
    \subfigure[Relative Network Growth]
    {\includegraphics[width=0.33\textwidth]{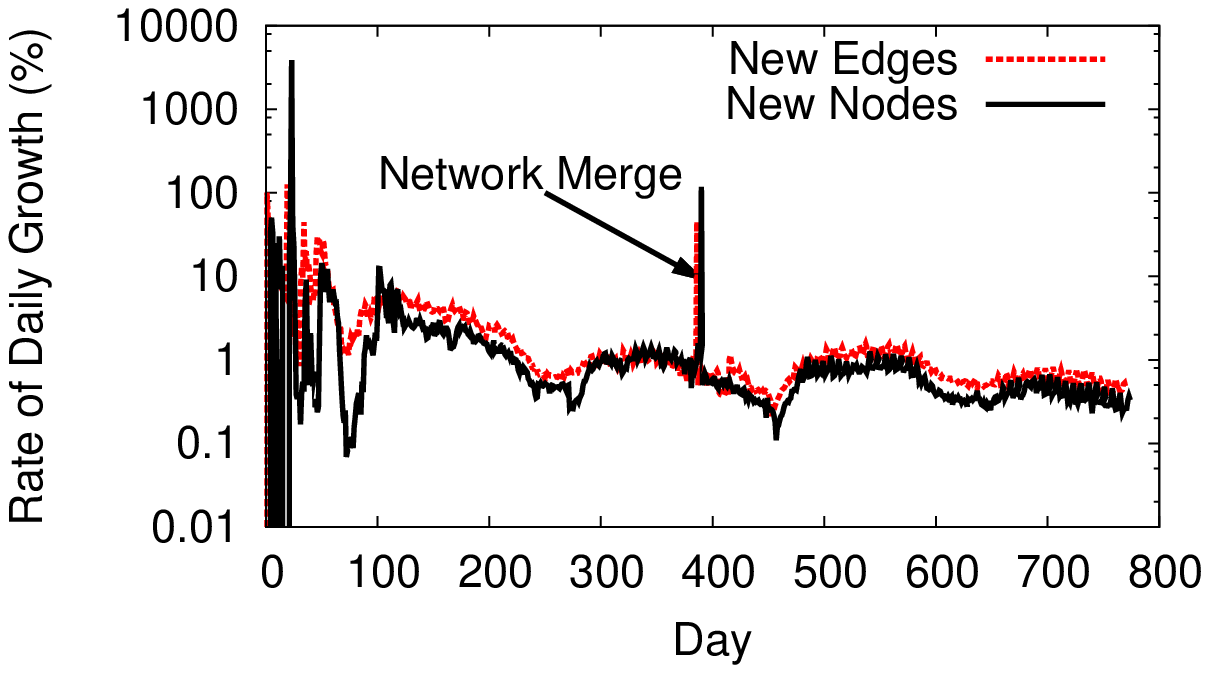}\label{fig:newper}}
    \subfigure[Average Node Degree]
    {\includegraphics[width=0.33\textwidth]{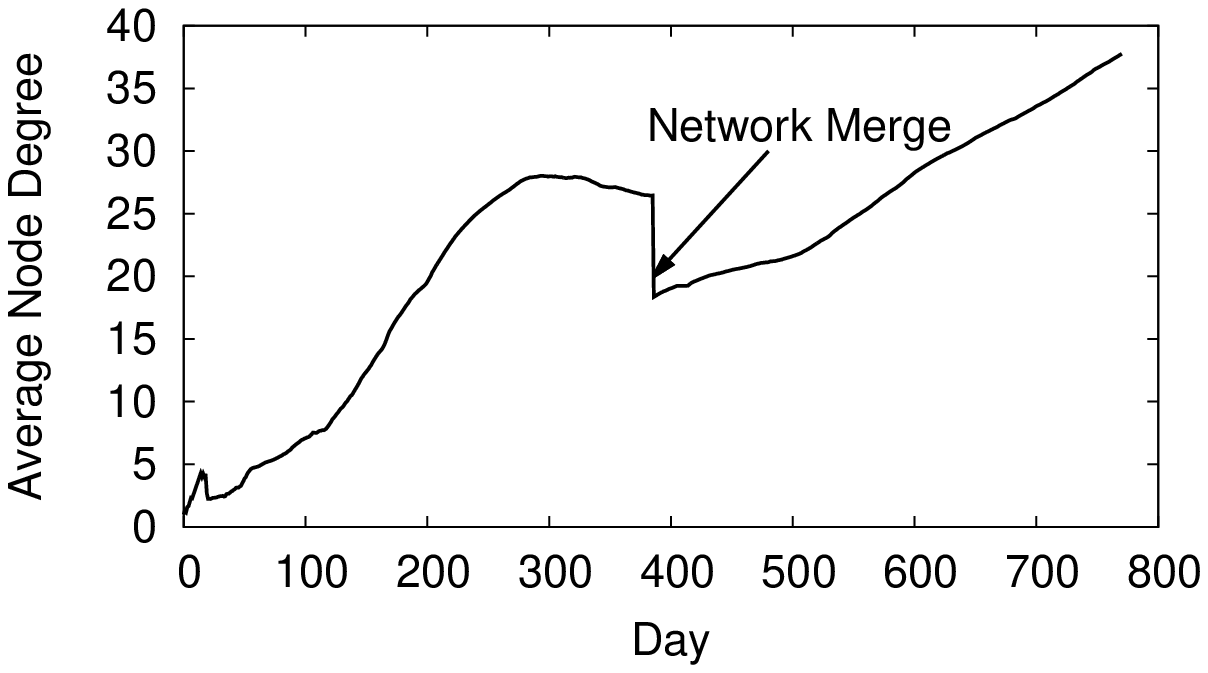}\label{fig:dyndeg}}
  }
  \mbox{ \subfigure[Average Path Length]
    {\includegraphics[width=0.33\textwidth]{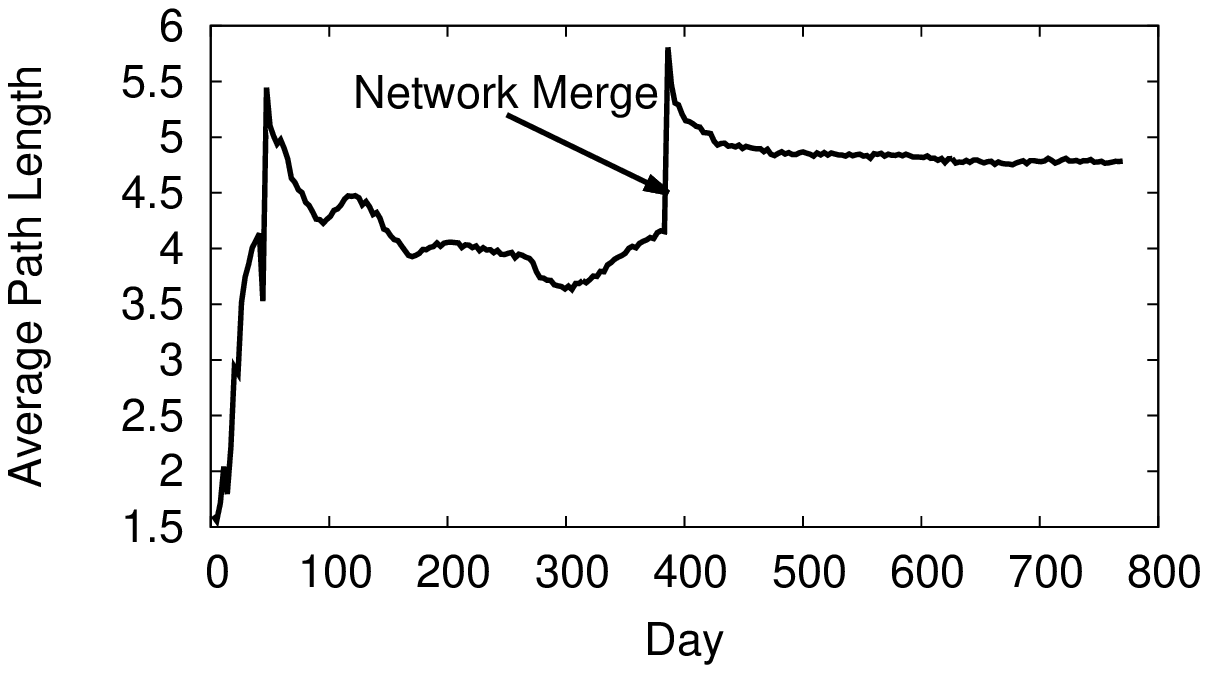}\label{fig:dynpath}}
\subfigure[Clustering Coefficient]
    {\includegraphics[width=0.33\textwidth]{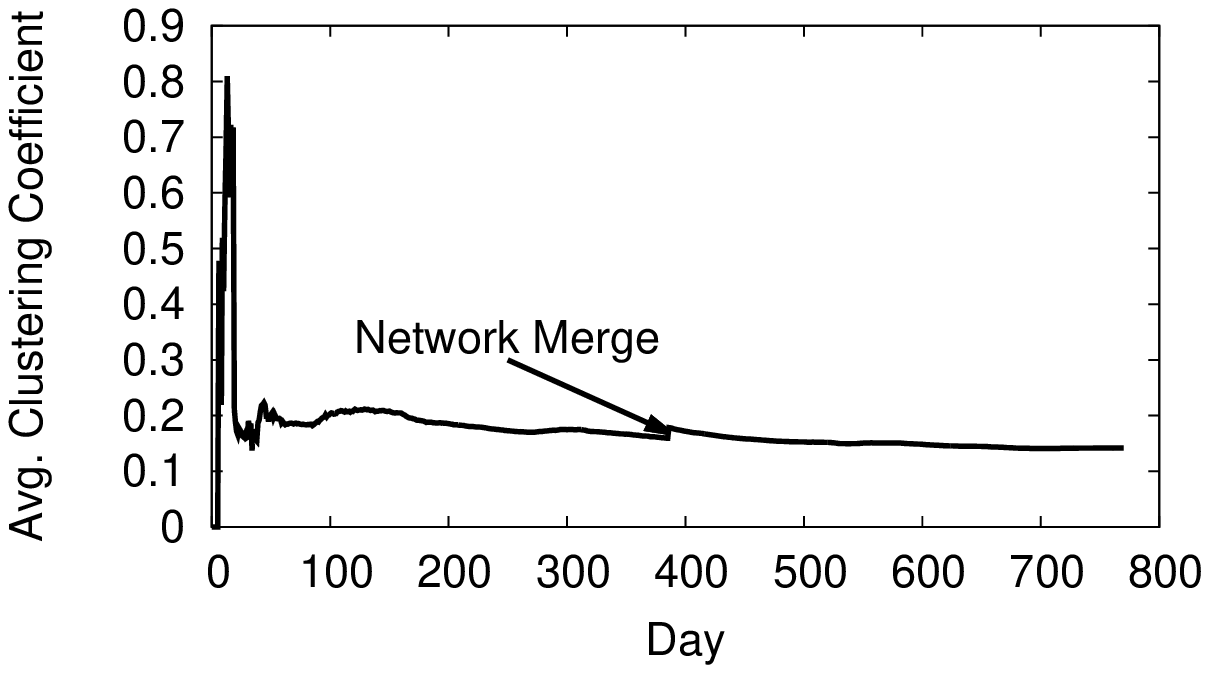}\label{fig:dyncc}}
    \subfigure[Assortativity]
    {\includegraphics[width=0.33\textwidth]{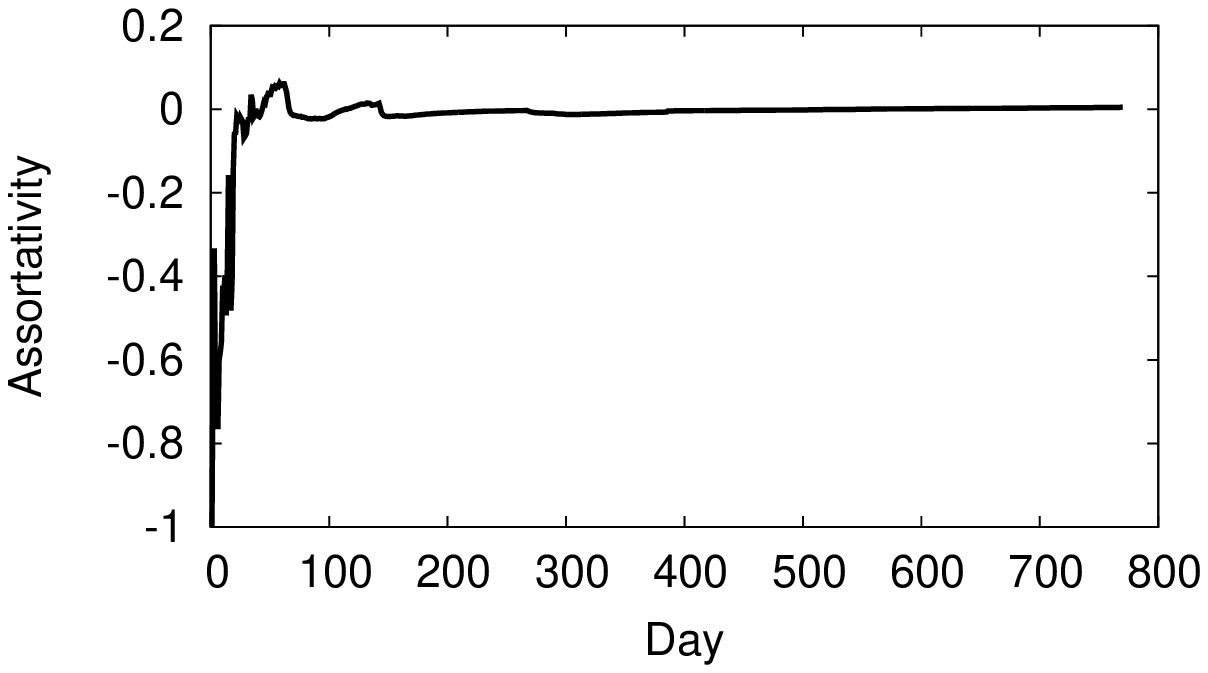}\label{fig:dynas}}
  } 
\caption{Network growth over time, and its impact on four important graph metrics.}
 \label{fig:network-results}
\end{figure*}

\section{Network Level Analysis}

We begin our study by first describing the dataset, and performing some basic
analysis to understand the impact of network dynamics on first order graph
metrics.  Our data is an anonymized stream of timestamped events shared with us by
Renren~\cite{renren-imc10}.  Our basic measurements in this section set the
context for the analysis of more detailed metrics in later sections.

\para{Renren Dynamic Dataset.}
The first edge in the Renren social network was created on November 21, 2005.
The network was originally named Xiaonei, or ``inside school,'' since it was
targeted as a communication tool for college students. Xiaonei expanded
beyond schools in November 2007, and changed its name to Renren (``
everyone'') in 2009.

Our anonymized dataset encompasses the timestamped creation events of all
users and edges in the Renren network. The dataset covers more than 2 years, starting on
November 21, 2005 and ending December 31, 2007.  In all, the dataset includes the
creation times of 19,413,375 nodes and 199,563,976 edges.  To perform
detailed analysis on the social graph, we produce 771 graphs
representing daily static snapshots from the timestamped event stream.  Note
that in this paper, we will use the term {\em node} to mean an OSN {\em user} and 
{\em edge} to mean a friendship link.

An unusual event happened on December 12, 2006, when Renren/Xiaonei
merged its social network with 5Q, a competing social network that
was created in April 2006.  On the merge date, Renren had 624K users with
8.2 million social links, and 5Q had 670K users with 3 million social
links.  Wherever possible, we treat the merge as an external event to
minimize its impact on our analysis of network growth. We present detailed
analysis of the network merge event in Section~\ref{sec:merge}.

On Renren, each user is limited to 1,000 friends by default. Users may pay a fee
in order to increase their friend cap to 2,000. However, prior work has shown that
very few users take advantage of this feature~\cite{renren-imc10}. We make the
same observation about our dataset: the number of users with $>$1,000 friends
is negligibly small.

\para{Network Growth.}
Figure~\ref{fig:new} depicts the growth of the Renren network
in terms of the number of nodes and edges added each day.  Day 0 is November
21, 2005.  Overall, the network grows exponentially, which is expected for a social
network. However, there are a number of real world events that temporarily slow
the growth, and manifest as visible artifacts in Figure~\ref{fig:new}.
The two week period starting at day 56 represents the Lunar New Year
holiday; a two-month period starting on day 222 accounts for summer vacation;
the merge with 5Q network causes a jump in nodes and edges on day 386;
additional dips for the lunar new year and summer break are visible
starting at days 432 and 587, respectively.  In Figure~\ref{fig:newper}, we
plot daily growth as a normalized ratio of network size from the previous
day.  It shows that relative growth fluctuates wildly when the network is
small, but stabilizes as rapid growth begins to keep rough pace with network
size.

\para{Graph Metrics Over Time.}
We now look at how four key graph metrics change over the lifetime of
our data stream, and use them to identify structural changes in the Renren
network.  We monitor average degree, average path length, average clustering
coefficient, and assortativity. As before, the analysis of each metric
starts from November 21, 2005.

\parab{Average Degree.} 
As shown in Figure~\ref{fig:dyndeg}, average node degree grows for much of
our observed time period, because the creation of edges between nodes
out paces the introduction of new users to the network.  This trend changes
around day 305, when a period of rapid growth in users starts to reduce
average degree.  This arises from a sudden influx of new users due to
several successful publicity campaigns by Renren.  In December 2006,
average degree drops suddenly when 670K loosely connected 5Q
nodes join the Renren network.  Average degree resumes steady growth
following the event, again showing edge growth out pacing node growth and
increasing network densification~\cite{leskovec2005graphs}.

\parab{Average Path Length.}  We follow the standard practice of sampling
nodes to make path length computation tractable on our large social graphs.
We compute the average path length over a sample of 1000 nodes from the SCC
for each snapshot, and limit ourselves to computing the metric once every
three days.  As seen in Figure~\ref{fig:dynpath}, the results are intuitive:
path length drops as densification increases ({\em i.e.} node degree increases).
There is a significant jump when 5Q joins Renren on day 386, but resumes a
slow drop as densification continues after the merge.

\parab{Average Clustering Coefficient.}  Clustering coefficient is a
measure of local density, computed as the ratio of the existing edges between
the immediate neighbors of a node over the maximum number of edges possible
between them.  We plot average clustering coefficient in Figure~\ref{fig:dyncc}.
In early stages of network growth (before day 60), the network was very small and
contained a large number of small groups with loose connections between
them. Groups often formed local cliques or near-cliques, resulting in
high clustering coefficients across the network.  Once the network grows in size,
average clustering coefficient transitions to a smooth curve and decreases slowly.
The network merge
produces a small jump, since the 5Q network had many small clusters of 3 or 4
nodes that boosted average clustering coefficient.
  
\parab{Assortativity.} Finally, we plot assortativity in
Figure~\ref{fig:dynas}. Assortativity is the probability of a node to connect
to other nodes with similar degree, computed as the Pearson correlation coefficient
of degrees of all node pairs.  In the early stages of the network, the graph is
sparse and dominated by a small number of supernodes connecting to many 
leaf nodes. This produces a strong negative assortativity that fluctuates and
then evens out as the network stabilizes in structure.  Assortativity evens
out at around 0, meaning nodes in Renren have no discernible inclination to be
friends with nodes of similar or different degree.

\parab{Summary.}  We observe that the high-level structure of the Renren social network 
solidifies very quickly.  Several key properties stabilize after the
first 2 months, with others establishing a consistent trend after 100 days.
While the notable network merge with 5Q introduces significant changes
to network properties, the effects quickly fade with time and continued user
growth.

%% file: preferential.tex
\begin{figure*}[t]
\centering
\subfigure[Distribution of Edge Inter-arrival Times]
{\includegraphics[width=0.32\textwidth]{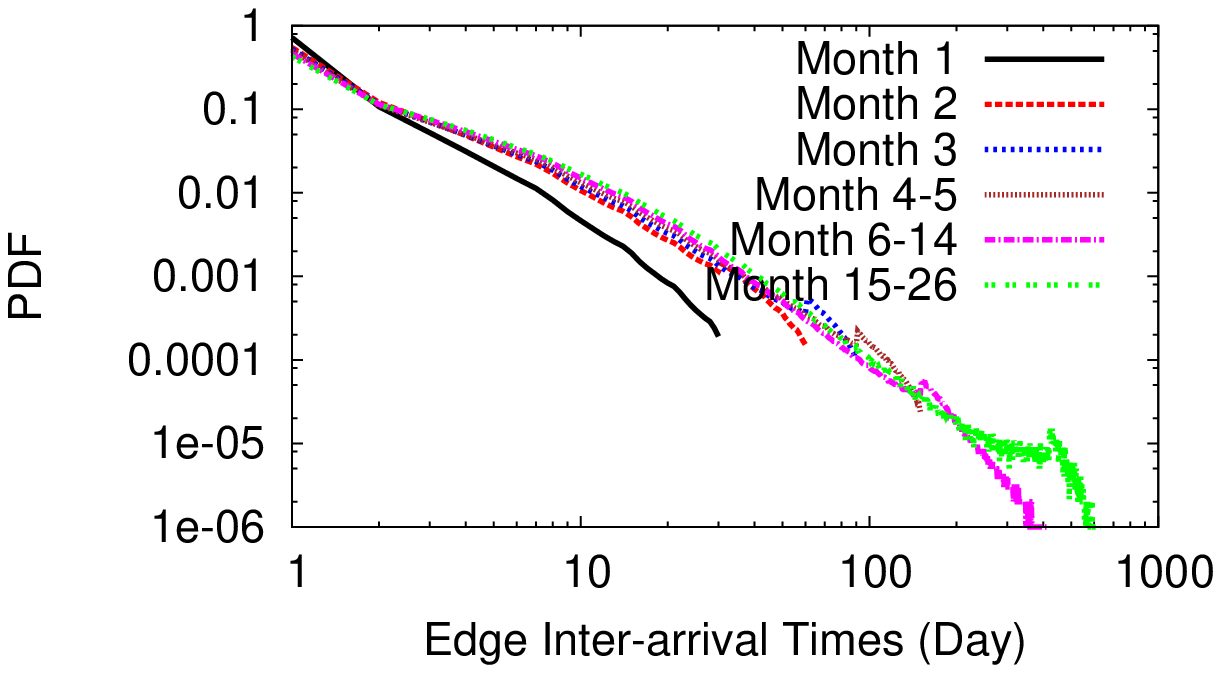}\label{fig:edgegap}}
\hfill
\subfigure[Edge Creation vs. Normalized Lifetime]
{\includegraphics[width=0.32\textwidth]{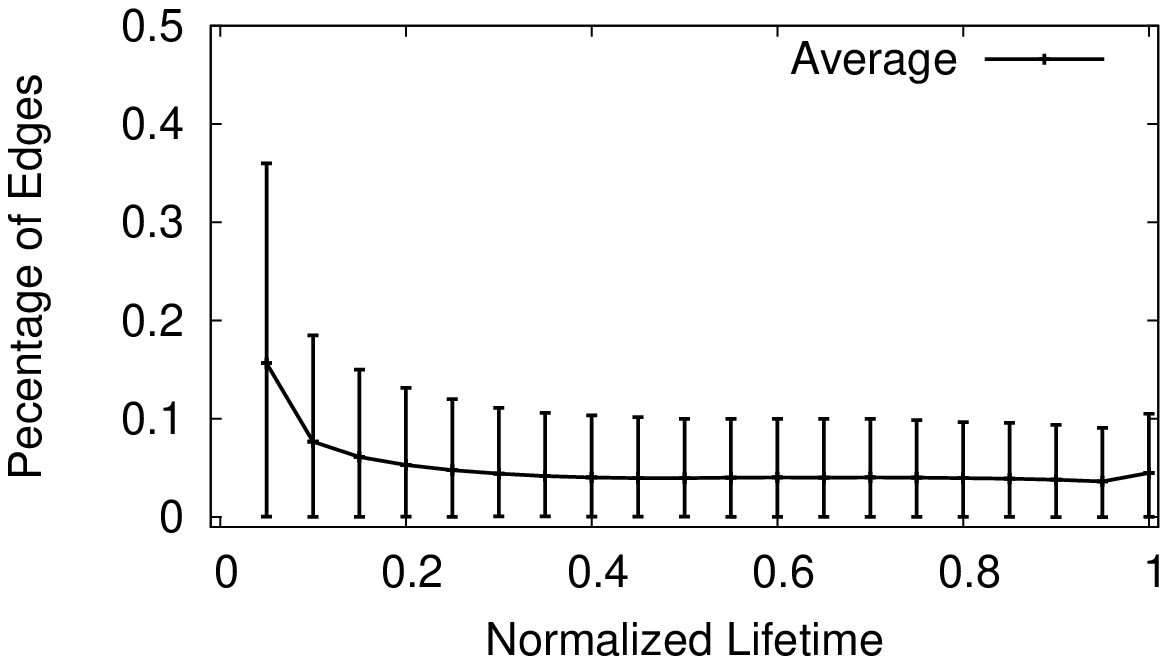}\label{fig:nodetrend}} 
\hfill
\subfigure[Node Age and Edge Creation]
{\includegraphics[width=0.32\textwidth]{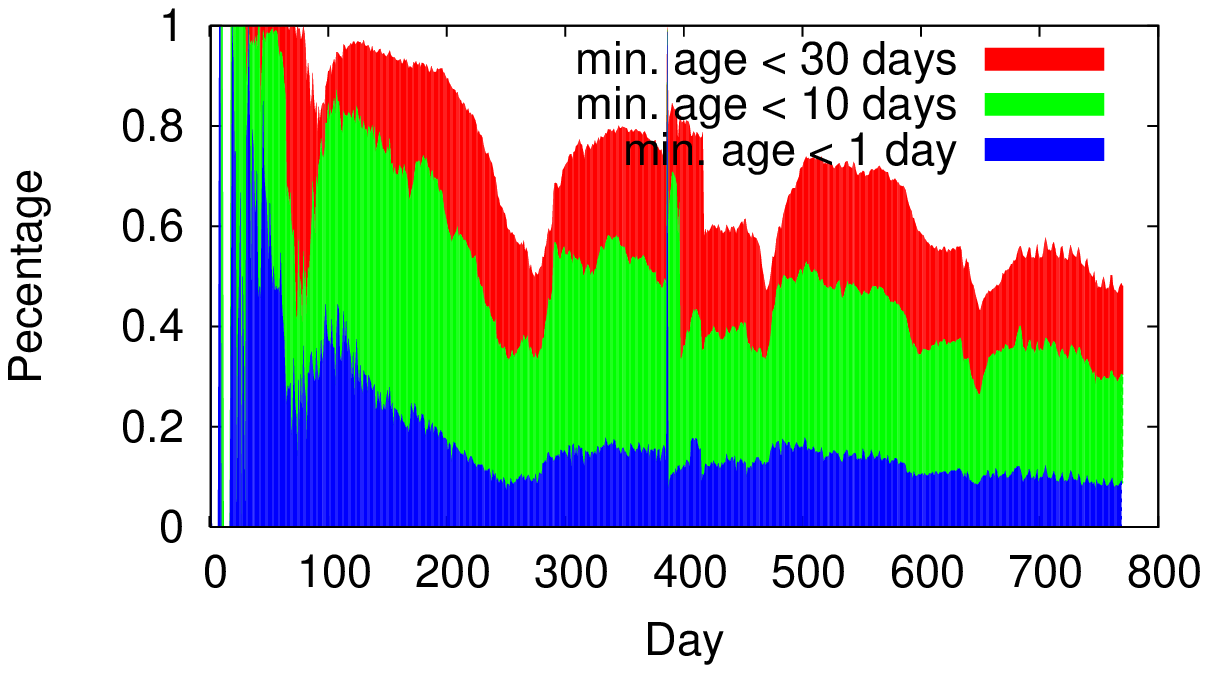}\label{fig:nodeage}}
\caption{Time dynamics of edge creation. (a) The probability distribution of the
  edge inter-arrival times follows a power-law distribution. (b) The normalized
  activity level over each
  user's lifetime. Users create most of her friendships early on. (c) The
portion of edges created by new nodes each day.  When the network is
young, new edges are mostly triggered by newly joined nodes. However, as the
network matures, the majority of new edges connect older users. 
}
 \label{fig:edgedelay}
\end{figure*}

\section{Edge Evolution}
\label{sec:preferential}

In this section, we study the behavior of individual nodes in terms of how
they build edges over time. Many studies have shown that nodes build edges following
the preferential attachment (PA)
model~\cite{barabasi1999emergence,leskovec2008microscopic,mislove2008growth,garg2009evolution}.
Specifically, when a new node joins the network and creates edges, it chooses
the destination of each edge proportionally to the destination's
degree. In other words, nodes with higher degrees are more likely to be
selected as the destination of new edges, leading to a ``rich get richer''
phenomenon.

Using the dynamic Renren network data, we extend the analysis of this model
in two new dimensions.  First, while PA defines how a sequence of edges is created
in logical order, we seek to understand how node activities correlate with
absolute time.  Second, we are interested in whether, as the network evolves,
the predictive ability of the PA model grows or weakens over time.

\subsection{Time Dynamics of Edge Creation}

\parab{Edge Inter-arrival.} We begin by analyzing the edge creation process
in absolute time, focusing on 
the speed that nodes add edges. First, we look at the inter-arrival time
between edge creation events.  For each node, we collect the inter-arrival
times between all its edges, then place them into buckets based on the
age of the node when the edge was created.  We then aggregate all users' data together
for each bucket, {\em e.g.} the ``Month 1'' bucket contains all edge
inter-arrival times where one or both of the nodes was less than 1
month old. 


We plot the results in Figure~\ref{fig:edgegap}. We observe that the time
gap between a node's edge creations follows a power-law distribution. The scaling
exponent is between 1.8 and 2.5 in Figure~\ref{fig:edgegap}. 
Overall, this power-law 
distribution provides a realistic model of a user's idle time between edge
creations at different stages of her lifetime.

\parab{Edge Creation Over Lifetime.} The above result motivates us to examine
the normalized activity level within each user's lifetime.  We plot in
Figure~\ref{fig:nodetrend} the distribution of new edges based on the
normalized age of the users involved.  To avoid statistical outliers, we
consider only nodes with at least 30 days of history in our dataset and
degree of at least 20.  As expected, users create most of their friendships
early on in their lifetimes.  Edge creation converges to a constant rate once
most of the offline friends have been found and linked.

\parab{Node Age and Edge Creation.}  We observe above that nodes tend to
generate a significant portion of their edges soon after joining the network.
Since most generative graph models use new nodes to drive edge creation, we
ask the question ``What portion of the new edges created in the network are
driven by the arrival of new nodes?''  For each day in our dataset, we take
each edge created on that day and determine its minimal age, {\em i.e.} the
minimum age of its two endpoints.  The distribution of this value shows what
portion of new edges are created by new nodes.

We compute and plot this distribution in
Figure~\ref{fig:nodeage}.  We show the relative contribution by nodes of different
ages by plotting three stacked percentages, showing the portion of daily new edges with minimal
age $\leq$ 1 day, $\leq 10$ days, and $\leq 30$ days.  We see
that when the network is young ($\leq 60$ days), the vast majority of new
edges connect brand new nodes ({\em i.e.} 1 day old).  As the network
stabilizes and matures, that portion quickly drops, and continues to decrease
over time.  Edges with minimal age of 10-30 days dominate new edges for much
of our trace, but their contribution steadily drops over time from
95\% around day 100 to 48\% by day 770.  Note that this drop occurs even
after the daily relative network growth has reached a constant
level (see Figure~\ref{fig:newper}). It is reasonable to assume that in
today's Renren network (4.5 years past the end of our data), the vast
majority of new edges connect mature users who have been in
the network for significant amounts of time.

\begin{figure*}[t]
\centering
\subfigure[$p_e(d)$, higher-degree node as destination]
{\includegraphics[width=0.32\textwidth]{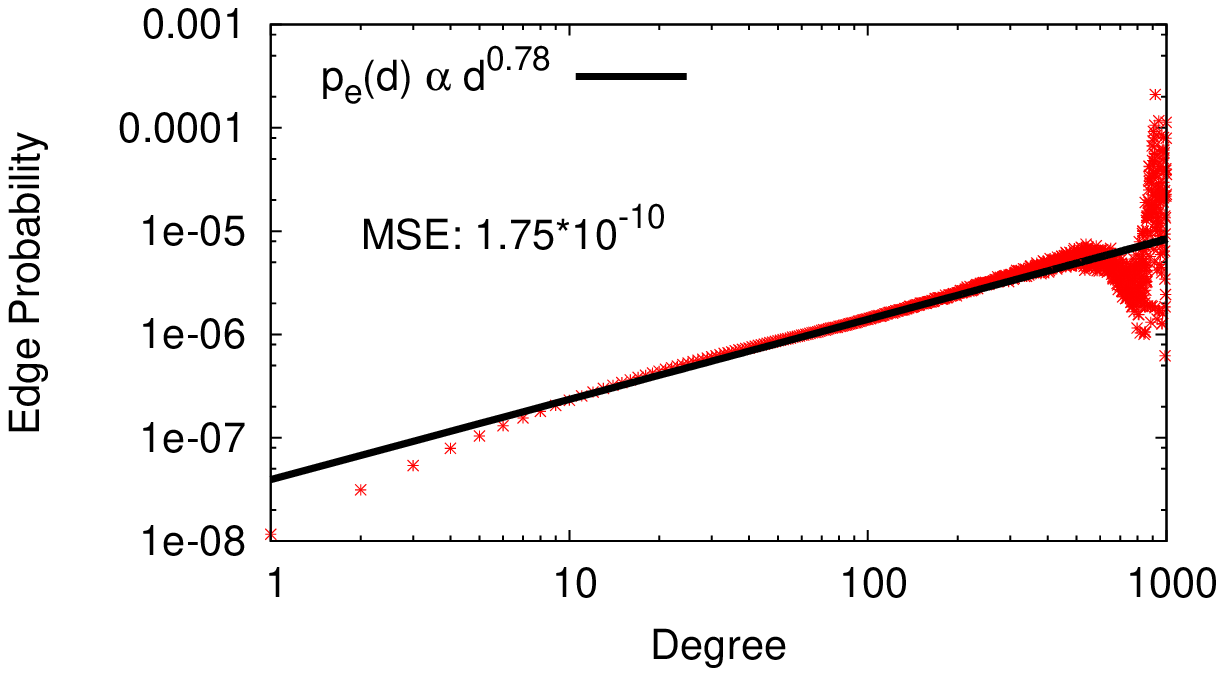}\label{fig:high}}
\hfill
\subfigure[$p_e(d)$, random destination selection]
{\includegraphics[width=0.32\textwidth]{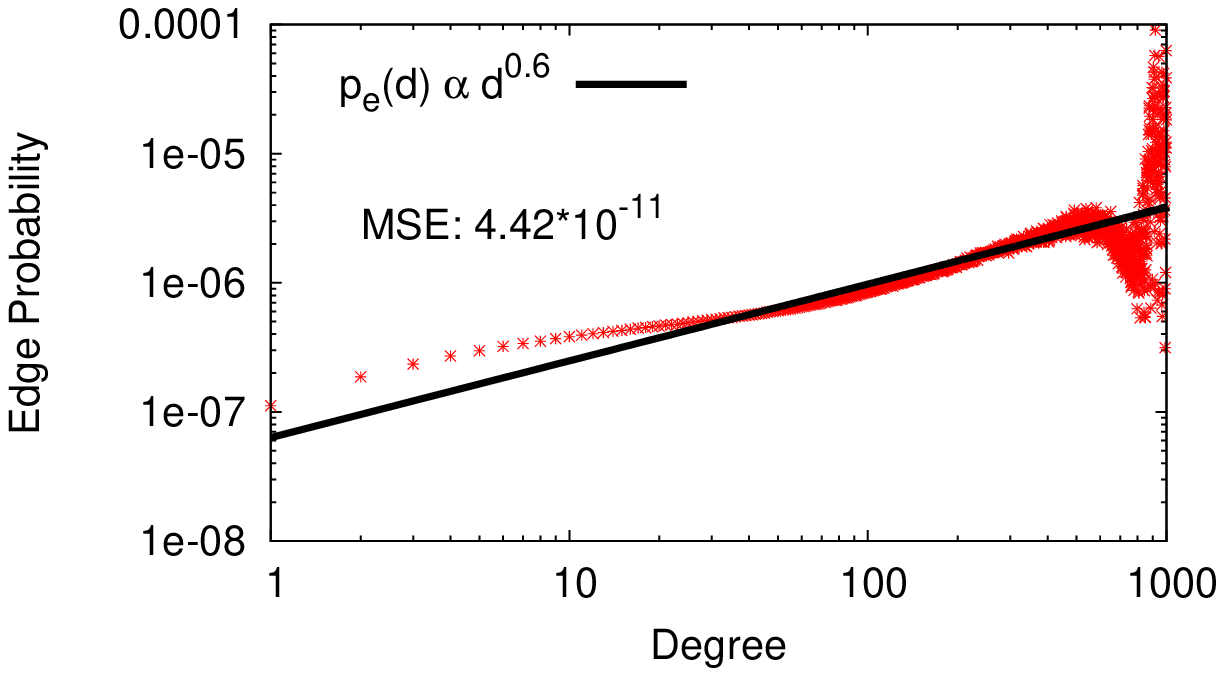}\label{fig:rand}} 
\hfill
\subfigure[Evolution of $\alpha(t)$]
{\includegraphics[width=0.32\textwidth]{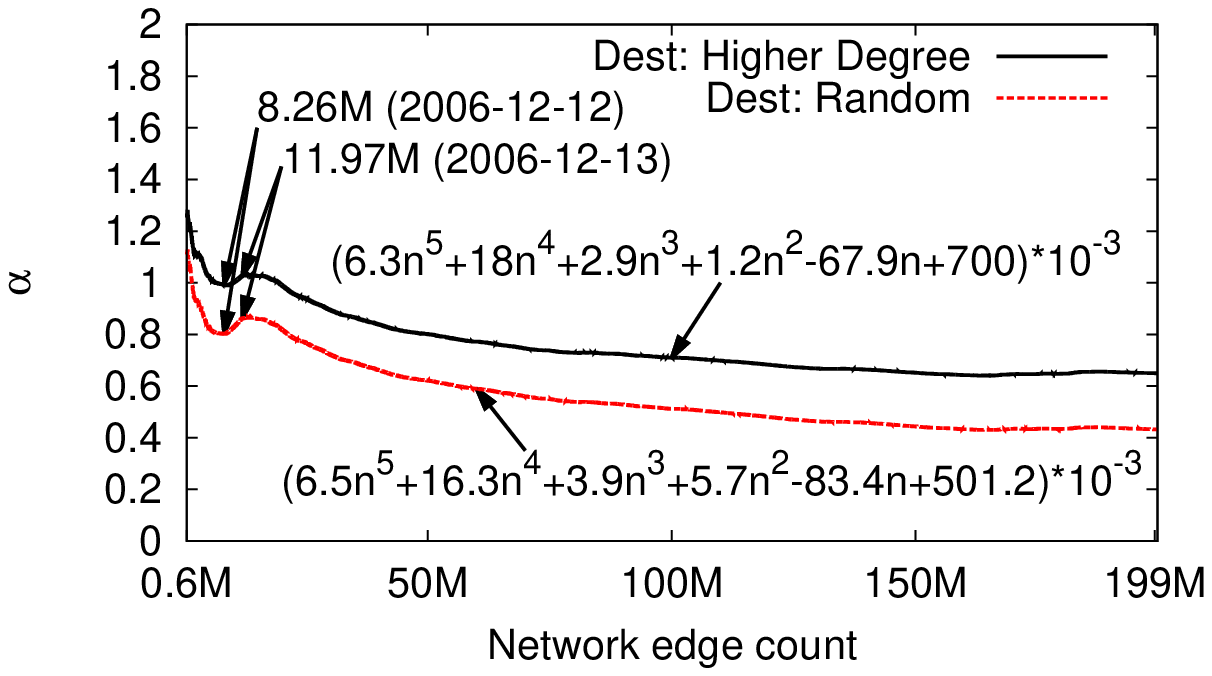}\label{fig:paslope}} 
\caption{(a)-(b) Fitting the measured edge probability $p_e(d)$ with
  $d^\alpha$, when the Renren network reaches 57M edges. In
  (a), $p_e(d)$ is calculated by selecting the higher-degree node as each
  edge's destination. In (b) the destination is selected randomly.  The mean
  square error (MSE) is very low, confirming the goodness of the fit.  (c)
  As the network grows, $\alpha$ drops from 1.25 to 0.65. It can be
 approximated by a polynomial function of the network edge count $n$.}
 \label{fig:paeg}
\end{figure*}

This result in Figure~\ref{fig:nodeage} is important, because it shows a
dramatic change in the driving force behind edge creation as the network
matures. Most generative graph models assume edge creation is driven by new nodes.
However, our data indicates that existing models will only accurately
capture the early stages of network creation.  Capturing the continuous
evolution of a mature network requires a model that not only recognizes the
contribution of mature nodes in edge creation, but also its continuous change
over time.

\subsection{Strength of Preferential Attachment} 
\label{sec:pa}   

Next, we take a look at the preferential attachment model and how well it
predicts changes over time and network growth.  We follow the method
in~\cite{leskovec2008microscopic} to measure the strength (or degree) of
preferential attachment using edge probability $p_e(d)$. This function defines the
probability that an edge chooses its destination with degree $d$, normalized
by the total number of nodes of degree $d$ before this time step:
\begin{equation}
p_e(d)=\frac{\Sigma_{t}{\{e_t(u,v) \wedge d_{t-1}(v)=d\}}}{\Sigma_{t}|v:d_{t-1}(v)=d|}
\label{eq:edgepb}
\end{equation}
where $\{e_t(u,v) \wedge d_{t-1}(v)=d\}=1$ if the destination $v$ of the edge
$e_t(u,v)$ is of degree $d$, and 0 otherwise. 

Intuitively, if a network grows following the PA model, its edge probability
$p_e(d)$ should have a linear relationship with $d$: 
$p_e(d)\propto d$. The authors of~\cite{leskovec2008microscopic} 
verified this conclusion using synthetic graphs,
and also tested the PA model on four real social networks: Flickr,
Delicious, Answers, and LinkedIn.  The first three networks follow the PA
model $p_e(d)\propto d^\alpha$ with $\alpha \approx 1$, while for LinkedIn,
$\alpha=0.6$.  From these observations, we can define a criterion for
detecting preferential attachment: when $\alpha \rightarrow 1$, the network
grows with a strong preferential attachment, and when $\alpha \rightarrow 0$,
the edge creation process becomes increasingly random.

Using this criterion, we validate the PA model over time on Renren by fitting
$p_e(d)$ measured at time $t$ to $d^{\alpha(t)}$ and examining $\alpha(t)$
over time.  Our study seeks to answer an important question: ``Does the
Renren network display the same level of preferential attachment consistently
over time?'' In other words, does $\alpha(t)$ stay constant over time?  And if
not, is the preferential attachment stronger (or weaker) at a particular
stage of network growth?


We make some small adjustments to the computation of $p_e(d)$ on the Renren data.
First, because our data does not state who initiated each friendship link (edge
directionality), we perform our test with two scenarios. The first is biased in
favor of preferential attachment because it always selects the higher degree end-point as
the destination. In the second scenario the destination is chosen randomly from the two
end-points.  Second, to make the computation tractable on our large number of
graph snapshots, we compute $p_e(d)$ once after every $5000$ new edges.
Finally, to ensure statistical significance, we start our analysis when the
network reaches a reasonable size, {\em e.g.} 600K edges.

\para{Results.} We start by examining whether $p_e(d) \propto
d^{\alpha(t)}$ is a good fit. For this we use the Mean Square Error (MSE)
between the measured $p_e(d)$ and the fitted curve.  We observe that the MSE
decreases with the edge count, ranging from 1.8e-5 to 3.5e-13. This
confirms that the fit is tight for the measured edge probability.  To
illustrate the results, Figures~\ref{fig:paeg}(a)-(b) show the edge
probability $p_e(d)$ when the network reaches 57M edges, using the two
destination selection methods. The corresponding MSEs of the fit are 1.7e-10
and 4.4e-11, respectively.




Next, we examine $\alpha(t)$ over time in
Figure~\ref{fig:paslope}.  We make two key observations. {\em First},
$\alpha(t)$ when using the higher-degree method is always larger than when
using random selection. This is as expected since the former is biased in favor of
preferential attachment. More importantly, the difference between the two
results is always 0.2. This means that despite the lack of edge destination information,
we can still accurately estimate $p_e(d)$ from these upper and lower bounds.  


{\em Second}, $\alpha(t)$ decays gradually over time, dropping from 1.25
(when Renren first launched) to 0.65 (two years later at 199M edges).
This means that when the network is young, it grows with a strong
preferential attachment. However, as the network becomes larger, its edge creation
is no longer driven solely by popularity. Perhaps this observation can be
explained by the following intuition. When a social network first launches,
connecting with ``supernodes'' is a key factor driving friendship
requests.  But as the network grows, it becomes harder to locate supernodes
inside the massive network and their significance diminishes.

Finally, we observe a small ripple at the early stage of
the network growth, when $\alpha(t)$ experiences a surge on December 12, 2006
(8.26M edges).  This is due to the Renren/5Q merge event, which generated a
burst of new edges that produce a bump in $\alpha(t)$ for that single day.

\subsection{Summary of Observations}

Our analysis produces three conclusions:

\begin{packed_itemize}
\item {\em In a node's lifetime, edge creation rate is highest
    shortly after joining the network and decreases over time.}
\item {\em Edge creation in early stages of network growth is driven by new
    node arrivals, but this trend decreases significantly as the network matures.}
\item {\em While edge creation follows preferential attachment, the strength
    degrades gradually as the network expands and matures.}
\end{packed_itemize}

These results set the stage for the following hypothesis.  An accurate
model to capture the growth and evolution of today's social networks should
combine a preferential attachment component with a randomized attachment
component. The latter would provide a degree of freedom to capture
the gradual deviation from preferential attachment.

%% file: community.tex
\section{Community Evolution}
\label{sec:community}

In online social networks, communities are groups of users who are densely
connected with each other because of similar backgrounds, interests or
geographic locations.  Communities effectively capture ``neighborhoods'' in
the social network.  As a result, we believe they represent the best
abstraction with which to measure the influence of social neighborhoods on
user dynamics.  We ask the question, ``how do today's social network
communities influence their individual members in terms of edge creation
dynamics?''  

To answer our question, we must first develop a method to scalably identify
and track communities as they form, evolve, and dissolve in a dynamic
network.  There is ample prior work on community detection in static
graphs~\cite{newman2004finding,clauset2004finding,Wakita2007,blondel10008fast}.
More recent work has developed several algorithms for tracking dynamic
communities across consecutive graph
snapshots~\cite{kim2009particle,sun2007graphscope,lin2008facetnet,tantipathananandh2007framework,tantipathananandh2009constant}.
Some of these techniques are limited in scale by computational cost, others
require external information to locate communities across snapshots of the
network. 

In the remainder of this section, we describe our technique for
scalably identifying and tracking communities over time. We then present our
findings on community dynamics in Renren, including community formation,
dissolution, merging, and splitting.  Finally, we analyze community-level
dynamics and use our detected communities to quantify the correlation between
node and community-level dynamics.

\subsection{Tracking Communities over Time}
\label{sec:imp}


Tracking communities in the presence of network dynamics is a critical step in our
analysis of network dynamics at different scales.  Prior work proved that
dynamic community tracking is an NP-hard problem~\cite{tantipathananandh2007framework}.
Current dynamic community tracking
algorithms~\cite{kim2009particle,sun2007graphscope,lin2008facetnet,tantipathananandh2007framework,tantipathananandh2009constant,greene2010tracking}
are approximation algorithms that ``track'' a community over multiple
snapshots based on overlap with an incarnation in a previous snapshot.  In
this section, we briefly describe our mechanism, which is a modified version
of \cite{greene2010tracking} that provides tighter community tracking across
snapshots, using the incremental version of the Louvain
algorithm~\cite{blondel10008fast}.  At a high level, we use incremental
Louvain to detect and track communities over snapshots, and use community
similarity to determine when and how communities have evolved.



\begin{figure*}[t]
\centering
\subfigure[Modularity]
{\includegraphics[width=0.32\textwidth]{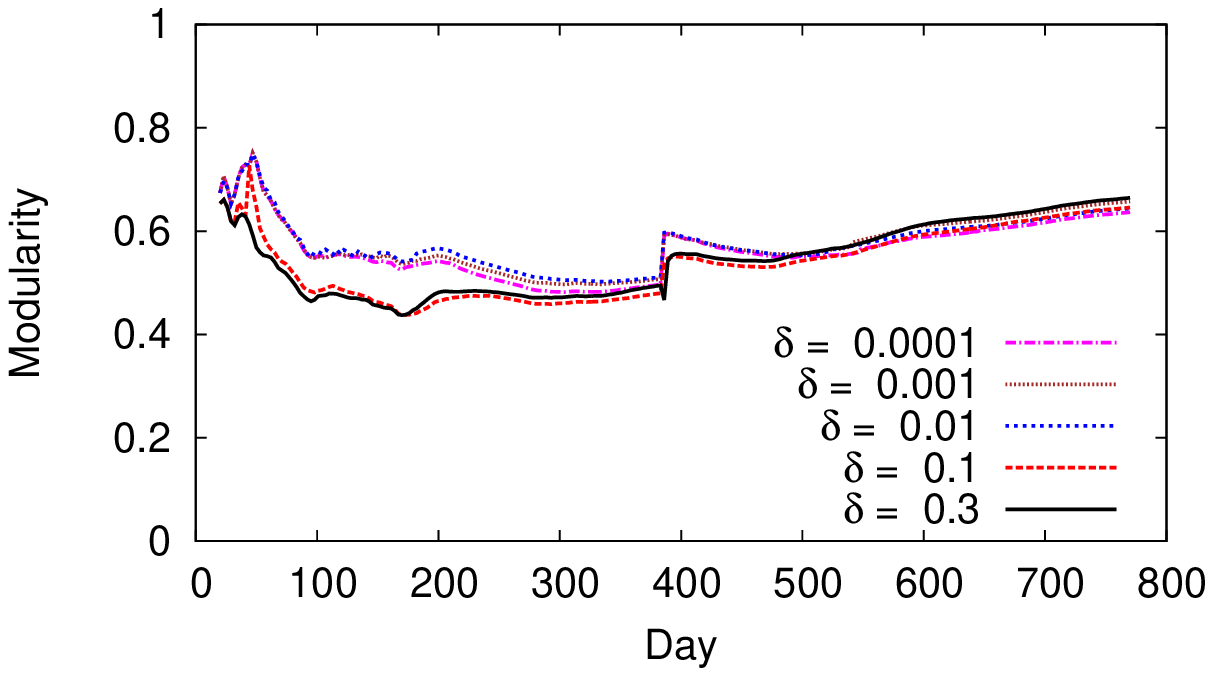}\label{fig:modl}}
\subfigure[Average Community Similarity]
{\includegraphics[width=0.32\textwidth]{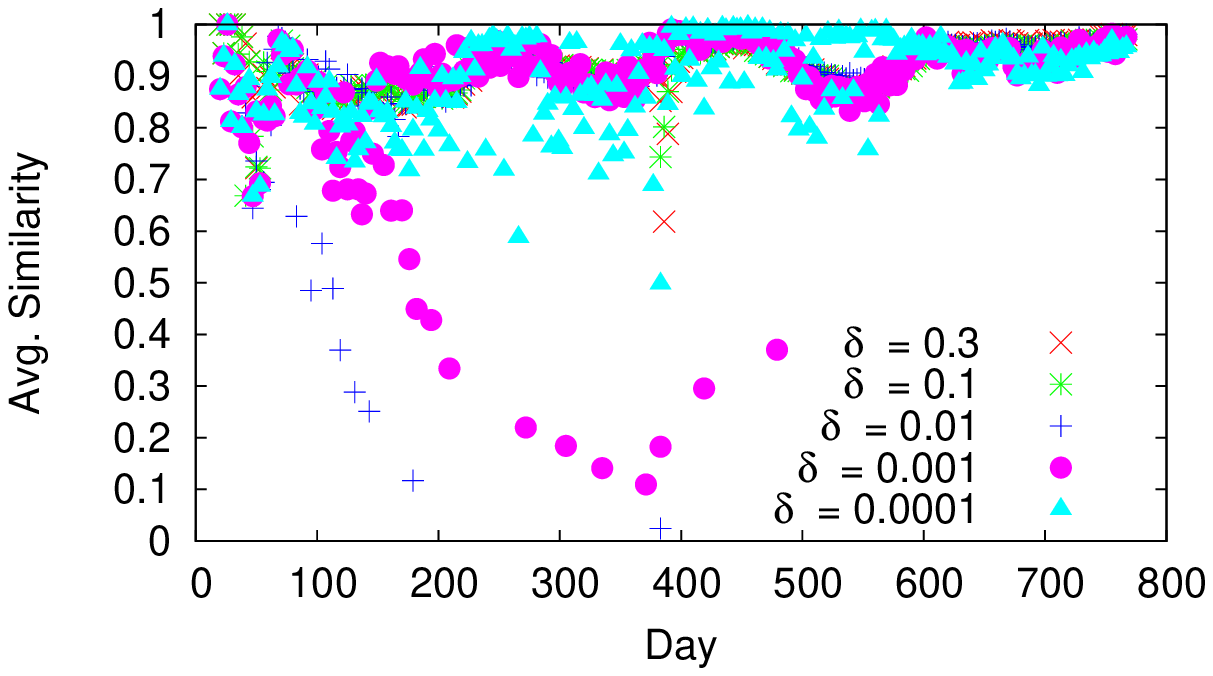}\label{fig:ac}}
\subfigure[Community Size Distribution on Day 602]
{\includegraphics[width=0.32\textwidth]{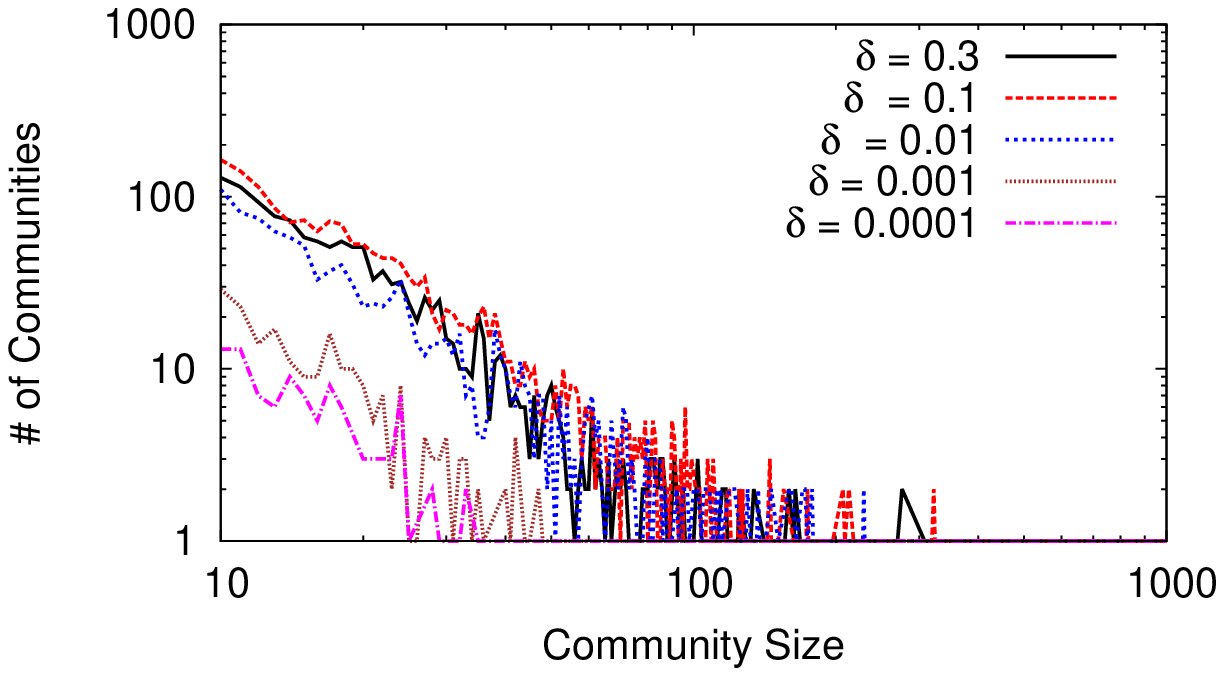}\label{fig:sizel}}
\caption{Tracking communities over time and the impact of $\delta$. (a) The value of modularity always
  stays above 0.4, indicating a strong community structure. The choice of
  $\delta$ has minimum impact, and $\delta=0.01$ is sensitive enough to
  detect communities.  (b) The value of average similarity
  over time at different $\delta$ values. Small $\delta$ values like 0.0001
  and 0.001 produce less robust results. (c) The distribution of community
  size observed on Day 602. The algorithm is insensitive to the choice of $\delta$ once
  $\delta\geq 0.01$. The same conclusion applies to other snapshots.}
 \label{fig:larget}
\end{figure*}

\para{Similarity-based Community Tracking.}
Louvain~\cite{blondel10008fast} is a scalable community detection algorithm
for static graphs based on optimizing modularity~\cite{newman2004finding}.
It uses a bottom up approach that iteratively groups nodes
and communities together, and migrates nodes between communities until the
improvement to modularity falls below a threshold $\delta$.  

Our approach leverages the fact that Louvain can be run in incremental mode,
where communities from the current snapshot are used to bootstrap the initial
assignments in the next snapshot.  Given how sensitive community
detection is to even small changes in modularity, this approach enables more
accurate tracking of communities by providing a strong explicit tie between
snapshots.  Finally, we follow the lead of \cite{greene2010tracking}, and
track communities over time by computing the similarity between communities.
Similarity is quantified as community overlap and is computed using set
intersection via the Jaccard coefficient.

\para{Community Evolution Events.} Using similarity to track communities
allows us to detect major community events, including their
birth, death, merges, and splits.  We define a community $A$ {\em splits} at
snapshot $i$ when $A$ is the highest correlated community to at least two
communities $B$ and $C$ at snapshot $i+1$.  When at least two communities $A$
and $B$ at snapshot $i$ contribute most of their nodes to community $C$ at
snapshot $i+1$, we say $A$ and $B$ have merged.  

When a community $A$ splits into multiple communities $X_1, X_2 ... X_n$,
we designate $X_j$ as the updated $A$ in the new snapshot, where $X_j$ is
the new community who shares the highest similarity with $A$.  We say that
all other communities in the set were ``born'' in the new snapshot.  Similarly,
if multiple communities merge into a single community $A$, we consider $A$ to
have evolved from the community that it shared the highest similarity
with.  All other communities are considered to have ``died'' in the snapshot.

\para{Choosing $\delta$.} The $\delta$ threshold in Louvain is an important
parameter that controls the trade off between quality of community detection
and sensitivity to dynamics. If $\delta$ is too small, the algorithm is too
sensitive, and over-optimizes to any changes in the network, needlessly
disrupting the tracking of communities.   If $\delta$ is too large, the process
terminates before it optimizes modularity, and it produces inaccurate
communities.

Choosing the best value for $\delta$ means optimizing for the dual metrics of
high modularity and robustness (insensitivity) to slight network
dynamics.  First, we use network-wide modularity as a measure of modularity
optimization for a given $\delta$ value.  Second, to capture robustness to
network dynamics, we use community similarity~\cite{greene2010tracking}:
the ratio of common nodes in two communities to the total number of different
nodes in both communities.  More specifically, for two consecutive snapshots,
we compute the average similarity between communities that exist in both
snapshots.  We run the Louvain algorithm on our snapshots using several different
$\delta$ threshold values, and select the best $\delta$ that generates both
good modularity and strong similarity.  We repeat this procedure on
shrinking ranges of $\delta$ until modularity and similarity can no longer be improved.

\para{Sensitivity Analysis.}
We run the Louvain algorithm on Renren dynamic graph snapshots generated every 3
days.  We start from Day 20, when the network is large enough (64 nodes) to support
communities, and only consider communities larger than 10 nodes to avoid small cliques.

We scale $\delta$ between 0.0001 and 0.3, and plot the resulting modularity
and average similarity in Figure~\ref{fig:larget}.  As shown in Figure~\ref{fig:modl},
in all snapshots the modularity for all thresholds is more than 0.4.
According to prior work~\cite{kwak2009mining}, modularity $\geq0.3$
indicates that Renren has significant community structure. As expected, a
threshold around 0.01 is sensitive enough for Louvain to produce communities
with good modularity.  Note that the big jump in
modularity on Day 386 is due to the network merge
event. 

Figure~\ref{fig:ac} shows that thresholds 0.0001 and 0.001 produce
lower values of average similarity ({\em i.e.} they are less robust and
more sensitive) compared to higher thresholds between 0.1 and 0.3. Thus,
Louvain with $\delta>0.01$ generates relatively good stability of
communities between snapshots.

Lastly, we examine whether detected communities are highly sensitive to
the choice of $\delta$. As an example, Figure~\ref{fig:sizel} plots the
distribution of community sizes observed on Day 602. The conclusion from
this figure is that once the threshold exceeds 0.01, the impact of $\delta$
on community size is reduced to a minimum. The same conclusion applies to
other snapshots as well.  



Based on the results in Figure~\ref{fig:larget}, we repeat the Louvain
algorithm within a finer threshold range of 0.01 to 0.1.  We find that a
threshold value of 0.04 provides the best balance between high modularity and
similarity.  We use $\delta = 0.04$ to track and measure dynamic communities
in the rest of our analysis on Renren.


\begin{figure*}[t]
\centering
\subfigure[Community Size Distribution] {\includegraphics[width=0.32\textwidth]{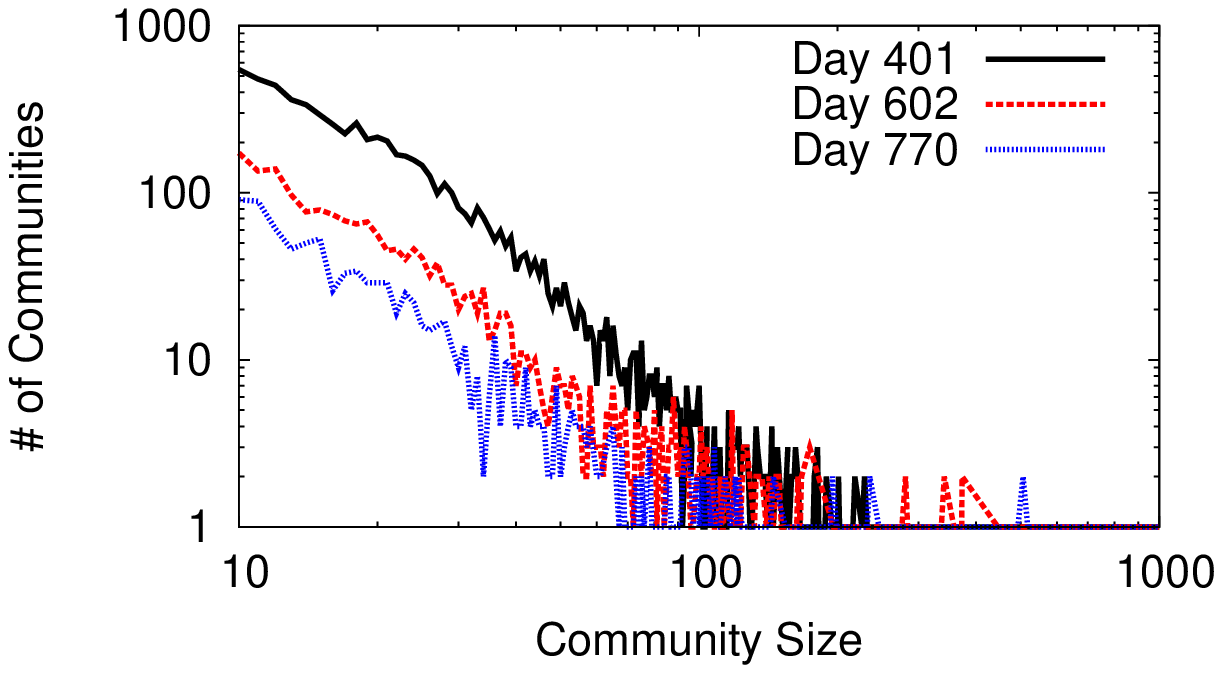}\label{fig:cn770}}
\subfigure[\% of Nodes Covered by Top 5 Communities]   
{\includegraphics[width=0.32\textwidth]{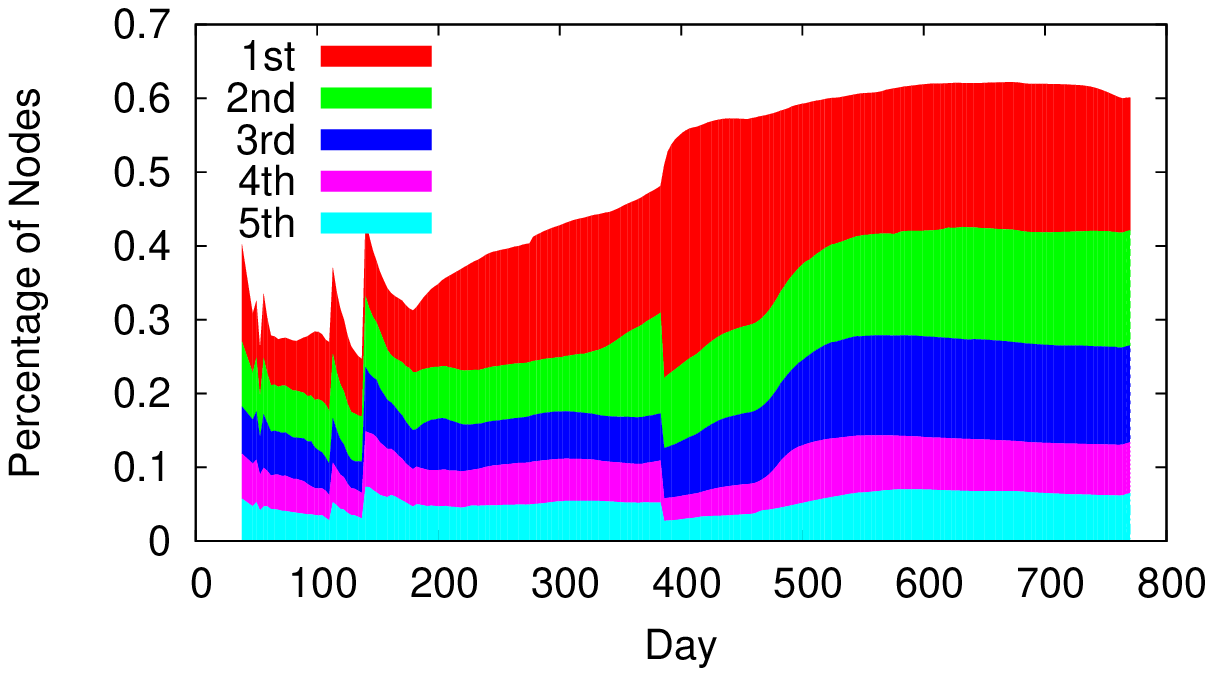}\label{fig:cnmedian}}
\subfigure[CDF of Community Lifetime] {\includegraphics[width=0.33\textwidth]{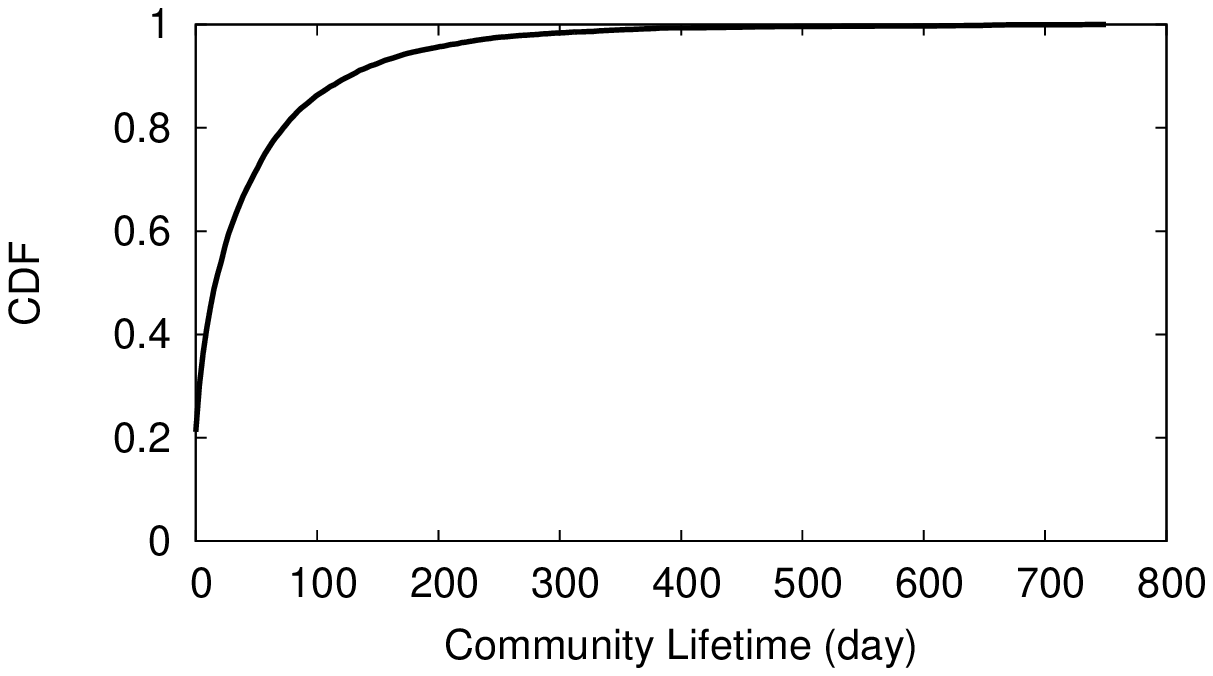}\label{fig:comlife}} 
\caption{Our analysis on the evolution of communities. (a) Community size
  distribution on Days 401, 602, and 770. All three lines follow a power-law
  distribution, and show a gradual trend towards larger communities. (b) The
  portion of nodes covered by the top 5 communities grows considerably as the
  network matures. (c) The distribution of community lifetimes. Most
  communities only stay in the network for a very short time, and are quickly
  merged into other communities. This indicates a high
  level of dynamics at the community level. } 
 \label{fig:comsize}
\end{figure*}

\subsection{Community Statistics Over Time}
\label{sec:dresult}
In the next three sections, we leverage our community tracking methodology to
analyze the dynamic properties of Renren communities.
We begin in this section by looking at the community size distribution,
how it changes over time, and the distribution of lifetimes for all communities.
In Section~\ref{subsec:merge}, we take a closer look at the dynamic processes
of community merges and splits.  We explore the possibility of predicting
community death from observed dynamics. Finally, in Section~\ref{sec:impact},
we analyze the impact of community membership on individual user dynamics, and
gauge how and to what extent community dynamics are observed to have influenced
individual user dynamics.


\para{Community Size.}
The size distribution of communities is an important property that reflects
the level of clustering in the network structure. Since the network structure
is constantly evolving, we can compute a community size distribution for each
snapshot in time.  We already observed in Figure~\ref{fig:sizel} that the
distribution of community sizes follows a power-law distribution. 

Our goal is to understand not only the instantaneous community size distribution, but also
how the distribution changes over time as the network evolves.
Thus, we compute the distributions for days 401, 602, and 770; 3 specific
snapshots roughly evenly spaced out in our dataset following the network
merge event.  We plot the resulting community size distributions in
Figure~\ref{fig:cn770}. The figure shows that the three snapshots consist of a large
number of small communities and a long tail of large communities, consistent
with the power-law distribution.  This is consistent
with other daily snapshots as well.  More importantly, these snapshots show a
gradual trend towards larger communities.  Over the year of time between
snapshots 401 and 770, the number of small communities shrunk by an order of
magnitude. In turn, the sizes of the largest communities increase significantly.

To take a closer at how communities grow over time, we focus on the
portion of the network that is covered by a small number of the largest
communities.  We take the top five communities sorted by size, and plot the
percentage of the overall network they contain in Figure~\ref{fig:cnmedian}.
We see that their
coverage of the network shows a clear and sustained growth over time.  They
grow from less than 30\% around day 100 to more than 60\% of the entire
network by the end of our dataset.  Over time, this trend seems to indicate
that as the network matures, connectivity becomes uniformly strong throughout
the main connected component, while distinctions between communities
fade.  


\para{Community Lifetime.}  In a dynamic network, how long a community
remains in the network is another important statistical property.  By using our
community identification method between snapshots, we measure the
distribution of community lifetime. Figure~\ref{fig:comlife} shows that most
of the communities only stay in the network for a very short period of time.
Specifically, 20\% of communities have lifetimes of less than a day,
meaning that they disappear in the next snapshot after they are first detected. 60\%
of the communities have lifetimes less than 30 days, at which point they are merged
into other communities. This shows an extremely high level of dynamics at the community
level.  

\subsection{Community Merging and Splitting}
\label{subsec:merge}
Community merging and splitting are the main reasons underlying community
death and birth.  Therefore, understanding these processes in detail is
critical to understanding dynamics at the community level as a whole.  We
study these processes in detail, with three questions in mind: What factors
influence the split and merge processes for communities?  What features, if
any, are good indicators for whether a community will merge soon?  Finally,
can we predict which communities will merge together?

\begin{figure*}[t]
\centering
\subfigure[Community Size Difference] {\includegraphics[width=0.31\textwidth]{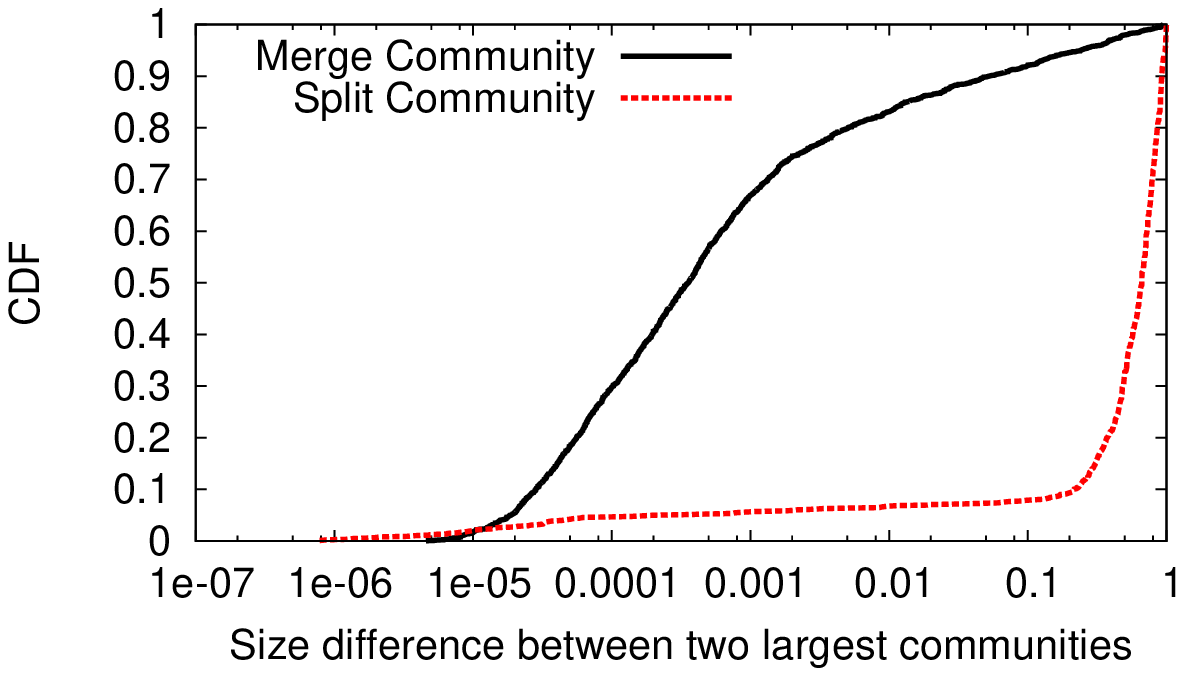}\label{fig:ratio}} 
\subfigure[Accuracy on Merge Prediction] {\includegraphics[width=0.31\textwidth]{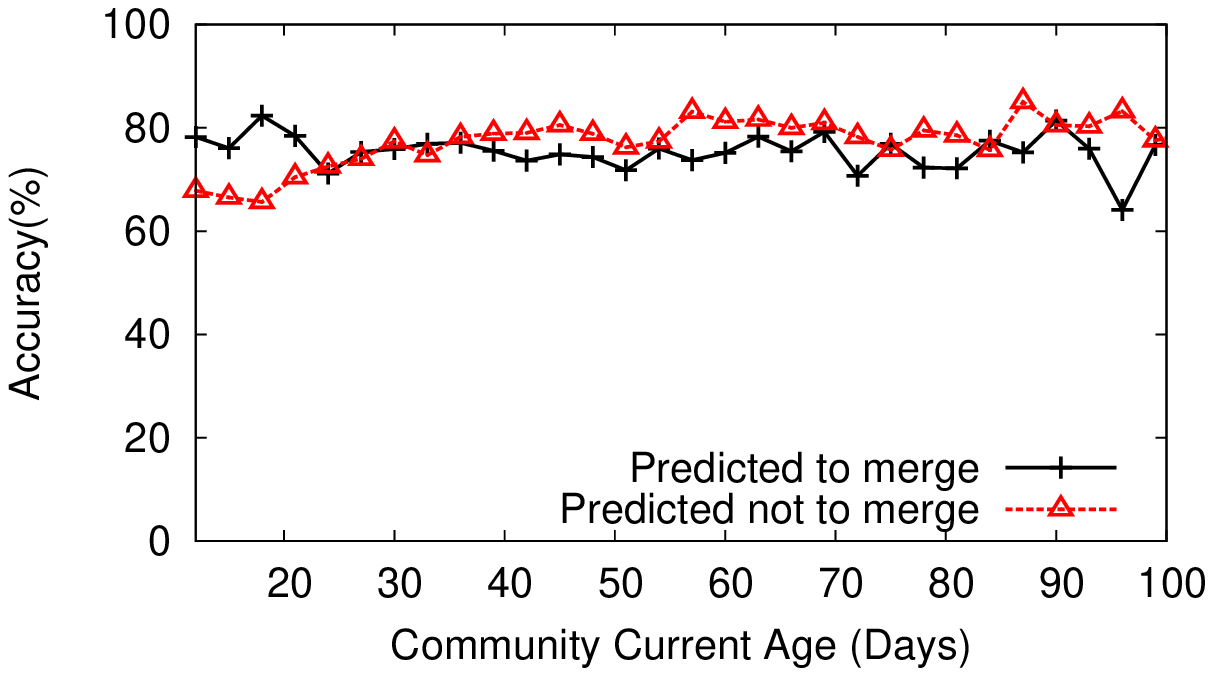}\label{fig:predic}} 
\subfigure[Patterns of Merged Communities] {\includegraphics[width=0.31\textwidth]{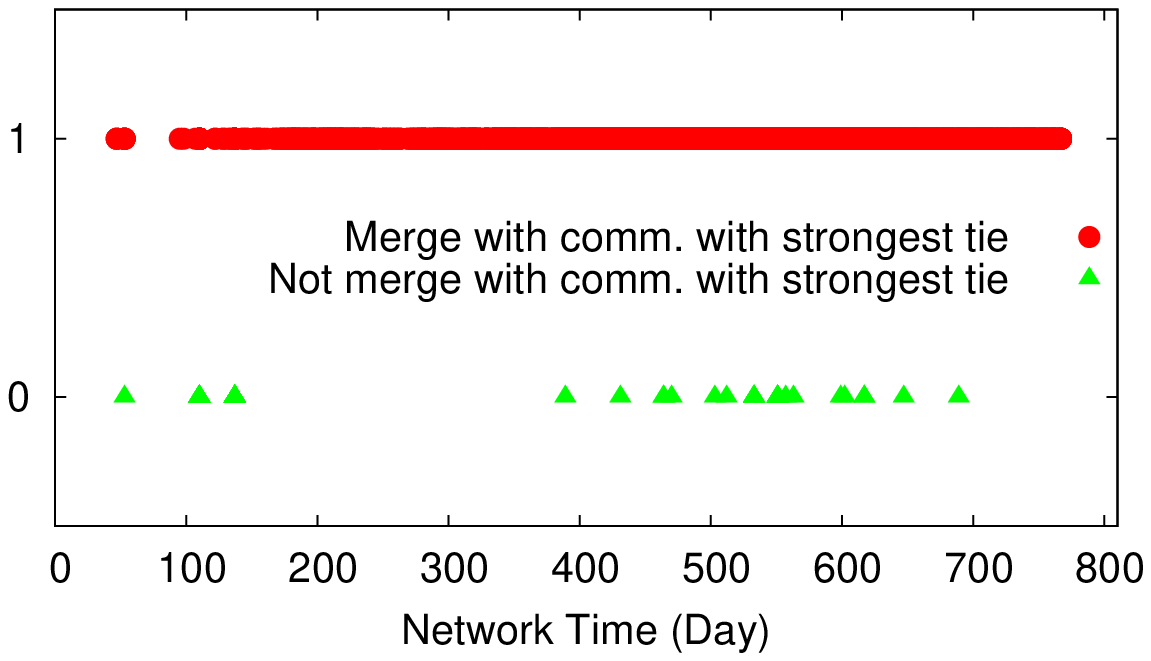}\label{fig:mergecom}}
\caption{Our analysis of community merging and splitting.  (a) The
  distribution of the normalized size difference between the largest two
  components when they split or merge. Small
  communities always merge into large communities, and a community tends to
  split into two communities of comparable sizes. (b)  The accuracy of our
  prediction on whether a community will merge with another in the next
  snapshot. We achieve a reasonably good accuracy of 75\%. (c) With very
  high probability (99\%), a community merges with the community that
  has the most edge connections (or the strongest tie) to itself.
 }
\label{fig:largethred}
\end{figure*}

First, we study whether community size impacts splitting or
merging.  For splitting events, we only consider the largest two
communities resulting from the split. Similarly for merge events, we focus on
the two largest communities merging to become one community.  We use as a
metric the ratio of the size of the second largest community to the size of
the largest community. The smaller the ratio is, the larger the
size difference is between the two communities.  In Figure~\ref{fig:ratio},
we plot the ratio of community splitting with a red line and community
merging with a black line. We observe that for 80\% of merged community
pairs, this ratio is less than 0.005. This reflects that for most merge
events, there is a large size discrepancy between the smaller
community and a larger community.  This is consistent with our observation
that small communities tend to disappear over time, while the biggest
communities continue to grow in size.  The community splitting process acts in
a totally different manner. The red line in Figure~\ref{fig:ratio} shows
that the ratio for 70\% split communities pairs are more than 0.5. Thus,
when a community splits into smaller communities, the community tends to
split into two comparable size communities.

\para{Predicting Merging.} Since community merge is the only
reason causing the death of the communities, we are curious whether there are
any structural features specific to the merge process, and whether
we can accurately predict if a community is going to merge with another in
the next snapshot.  We identify three structural metrics, including {\em
  community size}, {\em in-degree ratio}, the ratio of the edges inside a
community over the sum of the degrees of nodes in the community, and
the {\em similarity} of a community to itself in the previous snapshot
(defined in Section~\ref{sec:imp}).

Since these metrics evolve over time, we also consider short-term changes in
these features as additional factors.  For example, consider the community
size feature.  We can identify its {\em first order change indicator} as a
feature: if a community is smaller than its incarnation in the previous
snapshot, we use -1 to indicate the decrease. Similarly, we use 1 to mark an
increase and 0 to mark no change.  For each metric, we can also consider its
{\em second order change indicator}.  If the change in community size from
snapshot $i-1$ to $i$ is larger than the size change
from snapshot $i-2$ to $i-1$, we use 1 to indicate an acceleration in this
metric.  Similarly, we use -1 to mark a deceleration in this metric.  
In total, we start with the three basic metrics and add on their standard
deviation, their first order change indicator, and their second order change
indicator. 


Leveraging these feature metrics, we can now predict whether a community will
merge with another in the next snapshot. Specifically, we apply a Support
Vector Machines (SVM)~\cite{svm2001} over these features, together with the age
of each community.  For consistency, we do not consider communities created
on the day of the network merge with 5Q because those changes are driven by
external events.  To examine the accuracy of our prediction, we compute two metrics:
1) the ratio of the number of communities predicted to merge in the next snapshot to
the number of communities who actually merge, and 2) the ratio of the number of
communities predicted to not merge in the next snapshot to the number of
communities who do not merge.

Figure~\ref{fig:predic} plots our two accuracy metrics as a function of the
community age.  They show that our method
achieves reasonable prediction accuracy. It achieves an average accuracy of
75\% in predicting community merges and 77\% in predicting no
merges.  This means that we can reliably track communities' short-term
evolution.

We are also interested in predicting which destination community a given community
will merge into.  After examining each merged community pair, we make an
interesting observation. With a very high probability (99\%), a community $i$
will merge with another community $j$ that has the largest number of edges
to $i$, or the strongest tie with $i$. Figure~\ref{fig:mergecom} illustrates this trend
by plotting red dots for all merge events where a community merges with the peer with the
strongest tie, and a green triangle otherwise.  The results show that the trend is
consistent over time.  Thus, we conclude that the inter-community edge count is a reliable
metric for predicting the destination of community merges.

\input{impact}

\subsection{Summary of Results}
Our efforts on tracking and analyzing the evolution of communities lead to the
following key findings: 
\begin{packed_itemize}
\item {\em The Renren social network displays a strong community
    structure, and the size of the communities follows the power-law
    distribution.  }
\item {\em The majority of communities are short-lived, and within a few days
    they quickly merge into other larger communities.  The merges of these communities can
    be reliably predicted using structure features and dynamic metrics. }

\item {\em The membership to a community has significant influence on users'
    activity. Compared to stand-alone users, community users create edges more frequently,
    exhibit a longer
    lifetime, and tend to interact more with peers in the same community. }
\end{packed_itemize}

%% file: impact.tex
\begin{figure*}[tbph]
\centering
\subfigure[Edge Inter-arrival Time] {\includegraphics[width=0.33\textwidth]{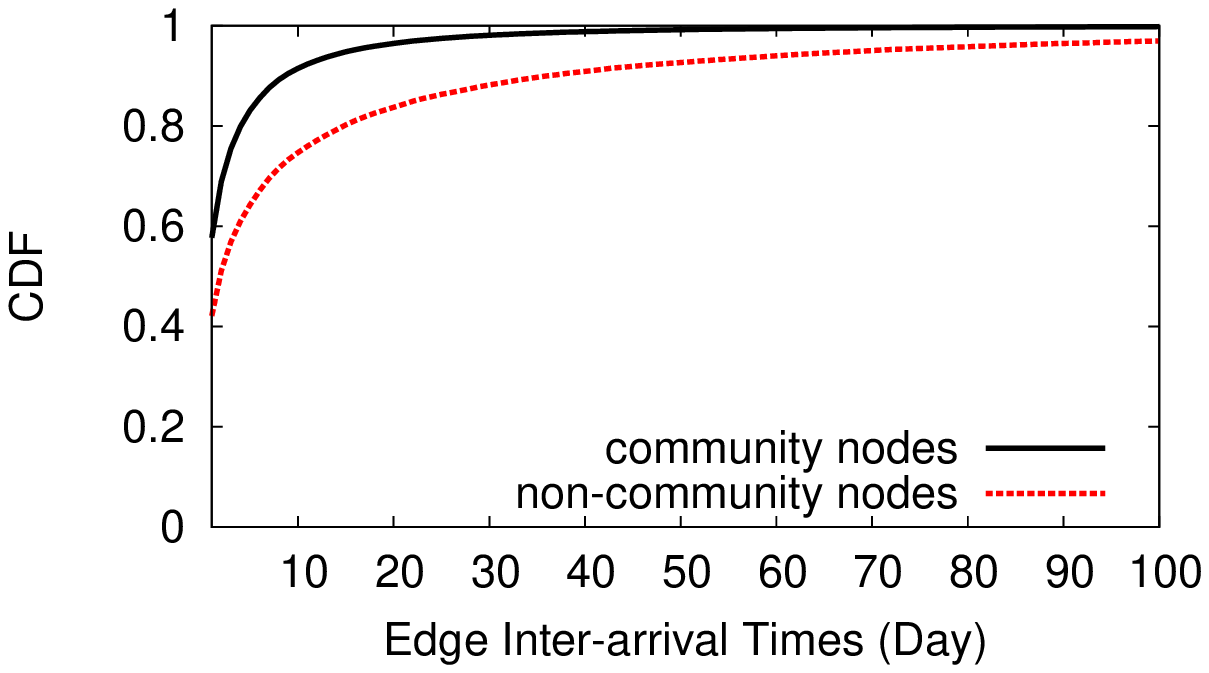}\label{fig:community-gap}}
\subfigure[Node Lifetime] {\includegraphics[width=0.33\textwidth]{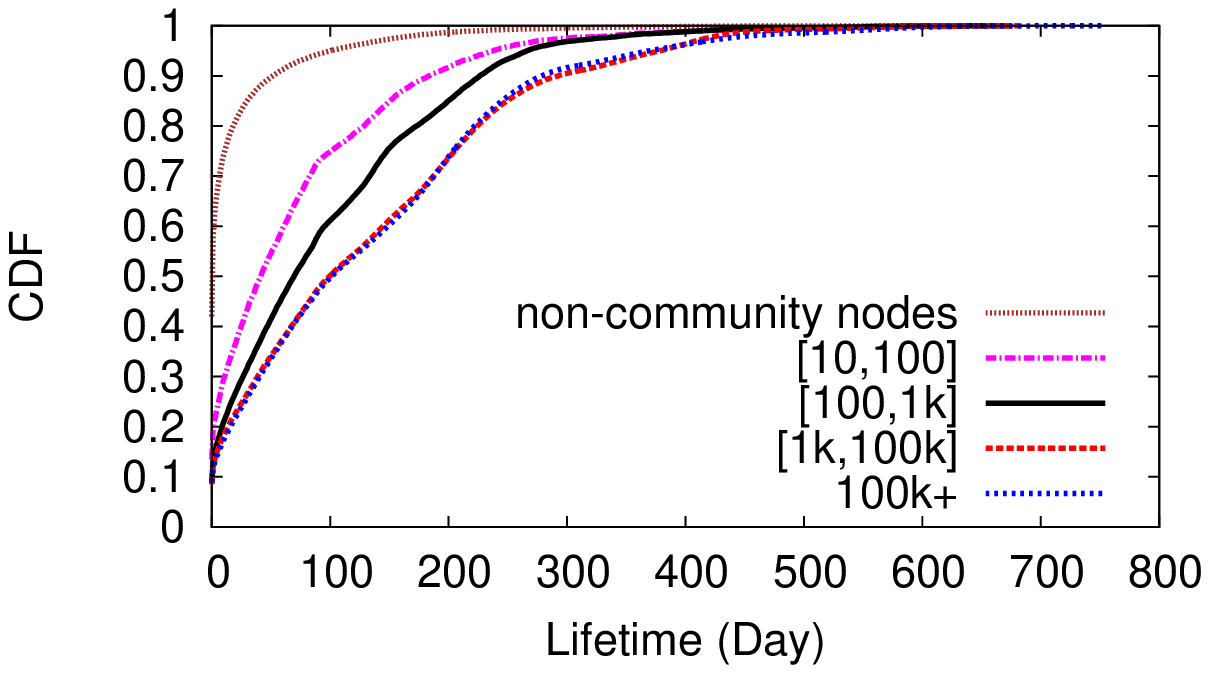}\label{fig:community-lifetime}}  
\subfigure[Edge In-Degree Ratio] {\includegraphics[width=0.33\textwidth]{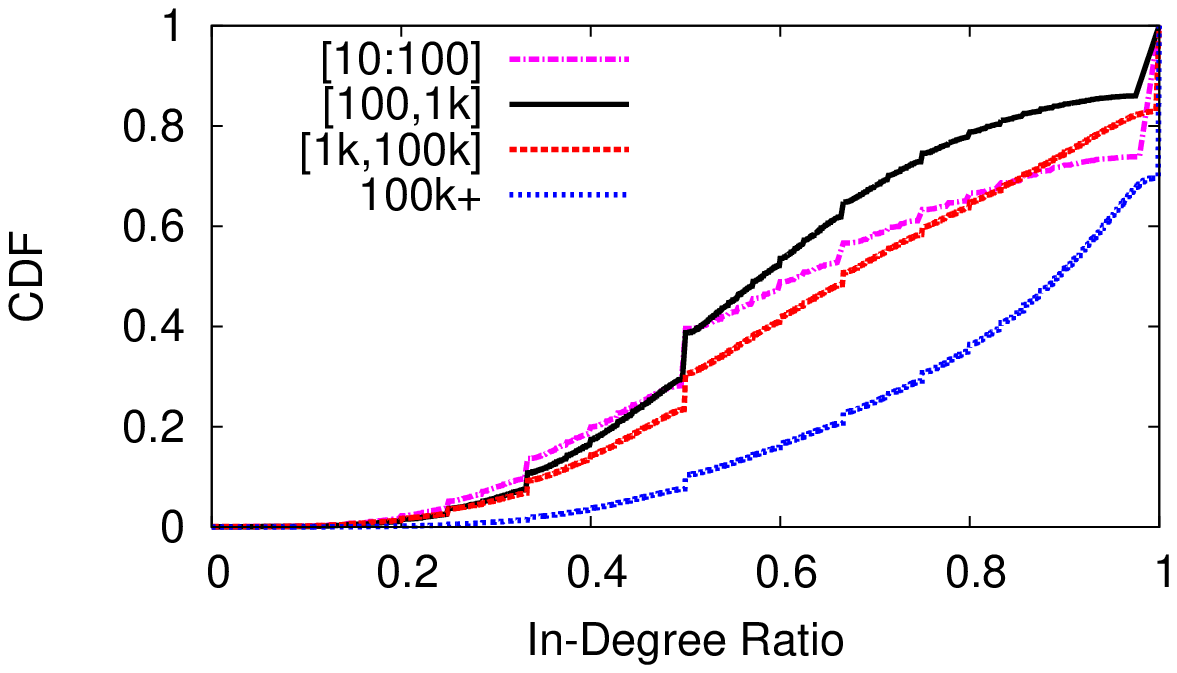}\label{fig:community-inoutratio}} 
\caption{Comparing activity of users inside and outside communities. Community
  users score higher on all dimensions of activity measures, confirming the
  positive influence of community on users.  (a) Edge inter-arrival time.
  Community nodes create edges more frequently than non-community
  nodes. (b) Node lifetime. Community users are grouped by their community
  sizes. $[x,y]$ represents communities of size between $x$ and $y$. Community nodes stay active longer than non-community  nodes. 
  (c) Community user's in-degree ratio.   Nodes in larger communities are
  more active within their own communities. }
\label{fig:community-impact}
\end{figure*}

\subsection{Impact of Community on Users}
\label{sec:impact}
To understand how communities impact users' activity,  we compare edge
creation behaviors of users inside communities
to those outside of any community. Overall, our results show that community
users score higher on all dimensions of activity measures, confirming the
positive influence of community on users. 


\para{Edge Inter-arrival Time.} Figure~\ref{fig:community-gap} plots the CDF of
edge inter-arrival times for community and non-community users.  We
observe that users within different communities display similar edge inter-arrival
statistics, and merge their results into a single CDF curve for clarity. The considerable
distance between the two curves confirms that community users are more enthusiastic in expanding their social
connections than non-community users. 

\para{User Lifetime.}  Next, we examine how long users stay active
after joining the network, and whether engagement in a community drives
up a user's activity span.  We define a user $i$'s lifetime as the gap between the
time $i$ builds her last edge and the time $i$ joins the
network.

Figure~\ref{fig:community-lifetime} plots the CDF of user lifetime for users
in different size communities as well as non-community users. $[x,y]$ represents
communities of size between $x$ and $y$. We find that the lifetime
distribution depends heavily on the size of the community. The larger the
community is, the longer its constituent user's lifetimes are.  Compared to non-community users,
users engaging in a community tend to stay active for a longer period of
time. This confirms the positive impact of community on users. 

\para{In-Degree Ratio.}   We also study how users within each community
connect to each other. We compute each user's in-degree ratio, {\em i.e.\/} the ratio of her edge count
within her community to her degree. Figure~\ref{fig:community-inoutratio} shows the CDF of the in-degree
ratio for users in communities of different sizes. 
We observe that users in larger communities have a larger in-degree ratio,
indicating that they form a greater percentage of edges within their own community.
In particular, 18-30\% of nodes only interact with peers in their
own communities, and the portion of these nodes grows with the community
size.  These results show that like offline communities, online social
communities also encourage users to interact ``locally'' with peers sharing
mutual interests.

%% file: netmerge.tex
\begin{figure*}[t]
\centering
\subfigure[Active users over time for Xiaonei] {\includegraphics[width=0.33\textwidth]{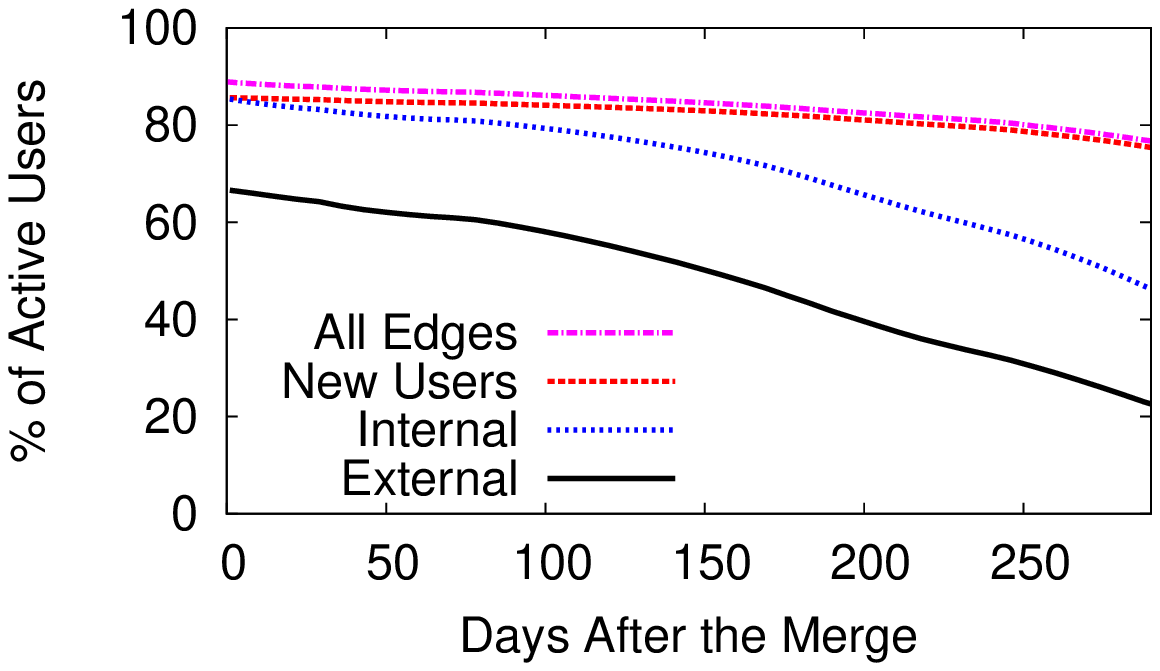}\label{fig:activenodesrr}} 
\subfigure[Active users over time for 5Q] {\includegraphics[width=0.33\textwidth]{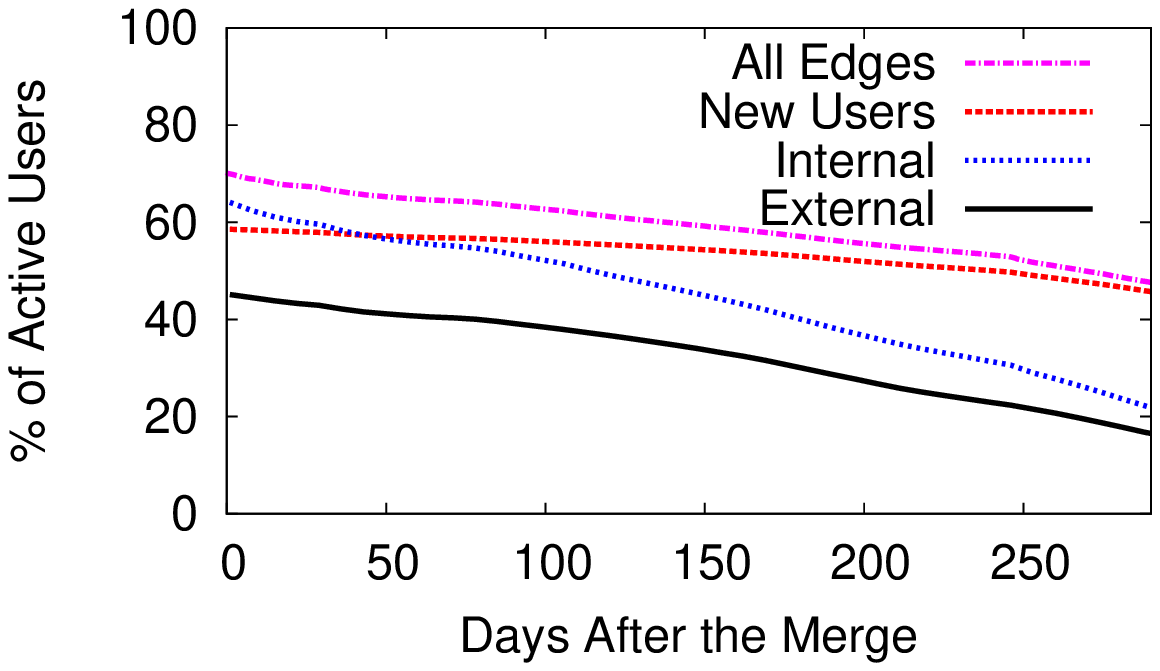}\label{fig:activenodes5q}} 
\subfigure[Edge creation per day] {\includegraphics[width=0.33\textwidth]{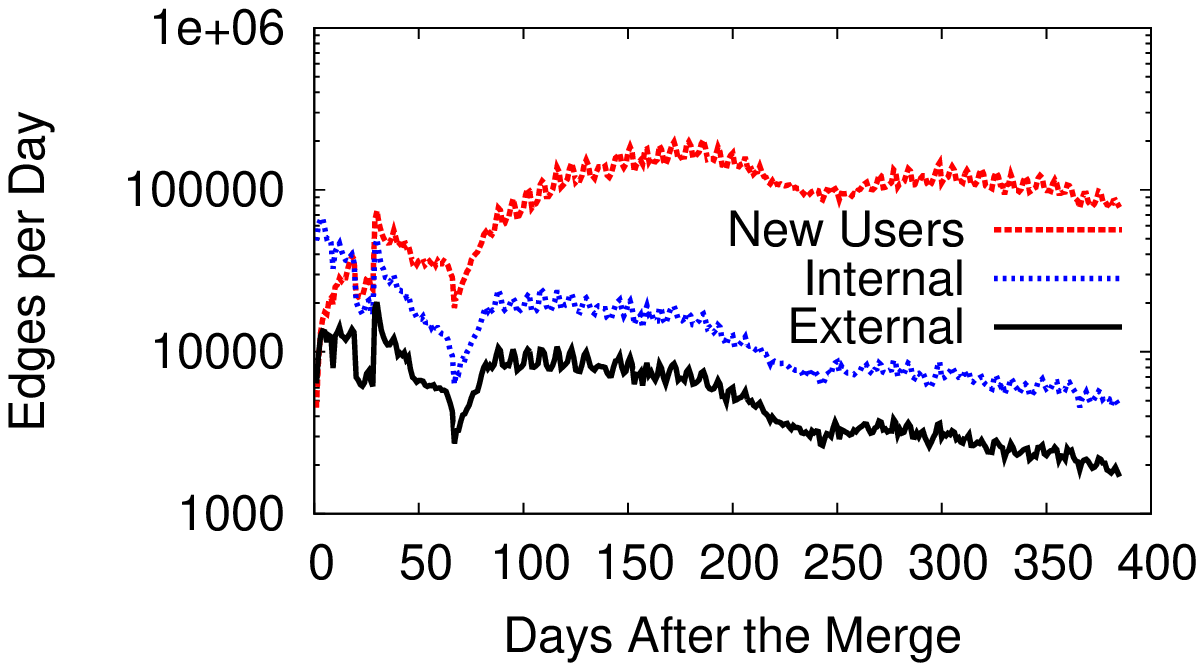}\label{fig:dayedge}} 
\caption{(a)-(b) The number of active users over time. Accounts that are inactive on day 0 after the merge are likely
to by discarded, duplicate accounts. Overall user activity declines over time. (c) Number of edges of different types
created per day after the merge. Edges to new users quickly become the most popular edge type, although there is
a small peak for external edges as well.}
\label{fig:mergeedges}
\end{figure*}

\section{Merging of Two OSNs}
\label{sec:merge}

On December 12, 2006 the OSN Xiaonei merged with another OSN called 5Q.
This combined entity became the Renren that exists today. Our access to the
graph topological and temporal data that characterizes this merge gives
us a unique opportunity to study how this network-level event impacts users' activity.

In this section, we analyze the forces at work during the merge.
First, we look at the edge creation activity of users over time in order
to isolate users that have become inactive. This enables us to estimate how
many duplicate accounts there were between Xiaonei and 5Q. Second, we examine
edge creation patterns within and between the two OSNs, and show that user
preferences vary by OSN and over time. We observe that the merge is the
primary driver of new edge creation for only a short time; edges to new users
that joined Renren after the merge rapidly take over as the driving force.
Finally, we calculate the distance between users in each group to quantify when
the two distinct OSNs become a single whole. We calculate that the average path
length from one OSN to the other drops rapidly in the days following the merge,
even when edges to new users are ignored. This demonstrates that the two OSNs
quickly become a single, indistinguishable whole.

\subsection{Background}

The predecessor to Renren, named Xiaonei, opened for business in
November 2005 to students in Chinese universities. 
Before the two networks merged, Xiaonei counted 624K active
users and 8.2M edges. 5Q was a competing OSN created in April 2006 that
also targeted university students. Before the merge, 5Q included 670K
active users and 3M edges.

On December 12, 2006, the two OSNs officially merged into a single OSN
known as Renren. During the merge, both OSNs were ``locked'' to prevent
modification by users, and all information from 5Q was imported and merged
into Xiaonei's databases. Starting the next day, users could log-in to the
combined system and send friend requests normally, {\em e.g.} users with
Xiaonei profiles could friend 5Q users, and vice versa. New users just joining
the system would not notice any difference between Xiaonei and 5Q user's profiles.

Since both 5Q and Xiaonei targeted university students, it was inevitable
that some users would have duplicate profiles after the merge. Renren allowed
users to choose which profile they wanted to keep, either Xiaonei or 5Q, 
during their first log-in to the site after the merge.

\para{Definitions.} In this section we investigate the details of the merge between
Xiaonei and 5Q. To facilitate this analysis, we classify the edges created
after the merge into three different groups. {\em External edges} connect
Xiaonei users to 5Q users, whereas {\em internal edges} connect users within
the same OSN. {\em New edges} connect a user in either OSN with a new user
who joined Renren after the merge. Time based measurements are presented
in ``days after the merge,'' {\em e.g.} one day after the merge is day 387
in absolute terms, since the merge occurs during day 386 of our dataset.

\subsection{Measuring the Merge}

\para{User Activity Over Time.} We start our analysis by examining
the number of active Xiaonei and 5Q users over time. We define a user as
``active'' if it has created an edge within the last $t$ days. In our
data, 99\% of Renren users create at least one edge every 94 days (on average),
hence we use that as our activity threshold $t$.

Figure~\ref{fig:activenodesrr} shows the number of active users over
time for Xiaonei, while Figure~\ref{fig:activenodes5q} focuses on the 5Q
users. Each ``all edges'' line highlights the number of users actively creating
edges in each group. Although we have 384 days of data after the merge,
the x-axis of Figures~\ref{fig:activenodesrr} and~\ref{fig:activenodes5q}
only extends 290 days. Since our minimum activity threshold is 94 days, we
cannot determine whether users have become inactive during the tail of our
dataset.

\begin{figure*}[t]
\centering
\subfigure[Ratio of the internal/external edges per day] {\includegraphics[width=0.33\textwidth]{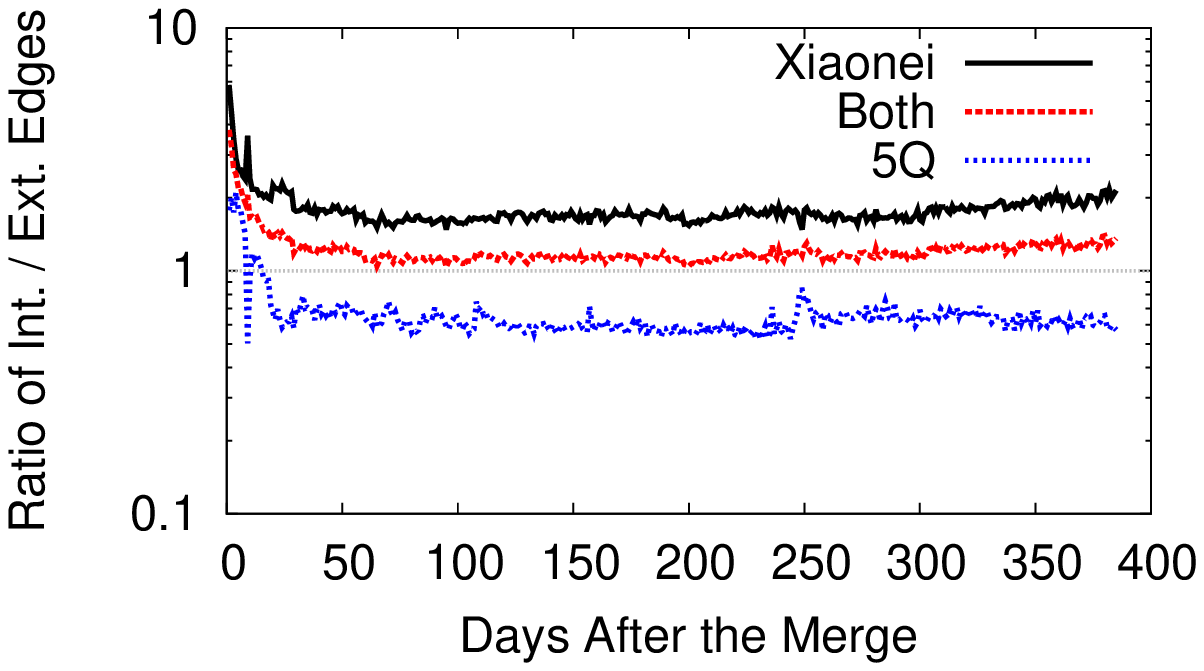}\label{fig:ratioive}}  
\subfigure[Ratio of new/external edges per day] {\includegraphics[width=0.33\textwidth]{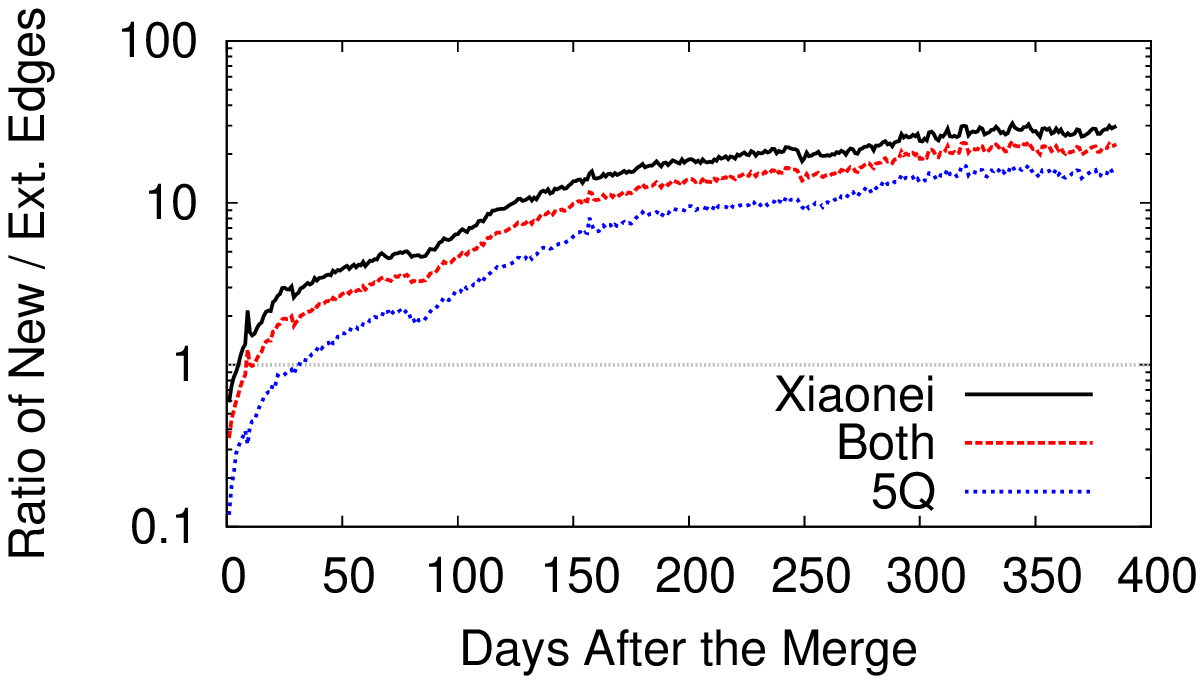}\label{fig:rationve}} 
\subfigure[Distance between the two OSNs over time] {\includegraphics[width=0.33\textwidth]{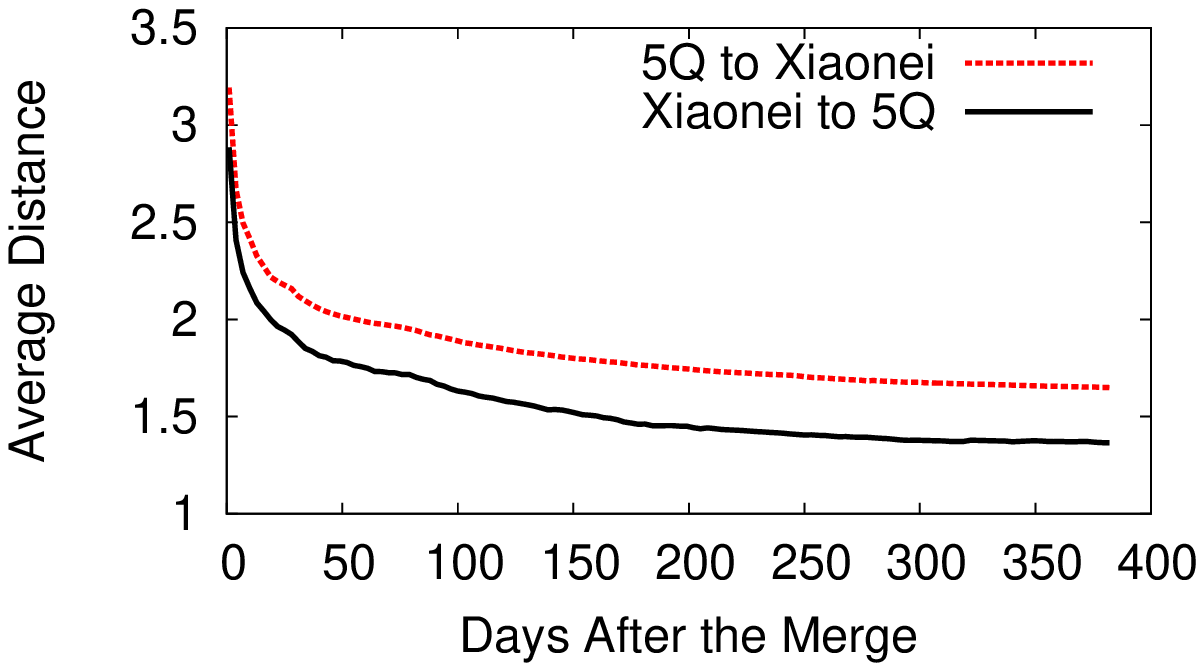}\label{fig:graphdistance}} 
\caption{(a) Ratio of internal to external edges over time. Xiaonei users create more edges overall, and are biased towards internal
edges, weighting the average upward. (b) Ratio of new to external edges per day. Both networks overwhelmingly prefer edges to new
users, although they reach this point at different rates. (c) The average distance in hops between the two OSNs drops over time as
more internal and external edges are created. By day 50, the two networks are essentially one large, well connected whole.}
\label{fig:edgeact}
\end{figure*}

We now address the question: {\em how many duplicate accounts were there
on Xiaonei and 5Q?} Users with accounts on both services
were prompted to choose one account or the other on their first log-in to Renren
after the merge. However, the discarded accounts were not deleted from the graph.
Thus, it is likely that any accounts that are inactive on the first day after the
merge are discarded, duplicate accounts.

Figures~\ref{fig:activenodesrr} and~\ref{fig:activenodes5q} reveal that 11\% of
Xiaonei accounts and 28\% of 5Q accounts are immediately inactive. Thus, it is
likely that at least 39\% of users had duplicate accounts on Xiaonei and 5Q
before the merge. Interestingly, users demonstrate a strong preference for
keeping Xiaonei accounts over 5Q accounts.

As time goes on, the number of active accounts in each group continues to drop.
Presumably, these users lose interest in Renren and stop generating new friend
relationships. After 284 days, the number of inactive Xiaonei accounts doubles
to 23\%, while on 5Q, 52\% of accounts are inactive. The relative decrease
in active accounts over time (12\% on Xiaonei versus 24\% on 5Q) demonstrates
that Xiaonei users are more committed to maintaining
their OSN presence. This observation corresponds to our earlier finding that users
with duplicate accounts tended to keep their Xiaonei accounts. Xiaonei users
form a self-select population of more active OSN users when compared to 5Q users.

The ``new users,'' ``internal,'' and ``external'' lines give the first glimpse
of the types of connections favored by Xiaonei and 5Q users. For each line, a user
is considered active only if they have created an edge of the corresponding type
in the last 94 days. Users in both graphs show similar preferences: edges to new
users are most popular, followed by internal and then external edges. The large
activity gap between internal and external edges highlights the strong homophily
among each group of users. Internal and external edge creation activity declines
more rapidly than edges to new users. This makes sense intuitively: the number of
Xiaonei and 5Q users is static, and hence the pool of possible friends slowly
empties over time as more edges are created. 

\para{Edge Creation Over Time.} Next, we switch focus to look at the characteristics
of edges, rather than individual users. By looking at the relative amounts of internal,
external, and edges to new users that are created each day, we can identify what types
of connections are driving the dynamic growth of Renren after the merge.

Figure~\ref{fig:dayedge} shows the number of internal, external, and new edges created
per day. Initially, internal and external edges are more numerous than edges to new users.
However, 3 days after the merge new edges begin to outnumber external edges, and by day
19 new edges out pace internal edges as well. This result demonstrates that new users
quickly become the primary driver of edge creation, as opposed to new edges
between older, established users. This is not surprising: since Renren is growing
exponentially, the number of new users eventually dwarfs the sizes of Xiaonei and 5Q,
which remain static.

Note that this result does not conflict with the results presented in
Section~\ref{sec:preferential}. Section~\ref{sec:preferential} examines the edge
creation patterns over the lifetime of {\em all} Renren users. In this section,
we are comparing the edge creation patterns of users who existed before the
merge to {\em everyone} who joined after. Thus, the age ``buckets'' in this section
are very course.

We now ask the question: {\em are there differences between the types of edges created
by Xiaonei and 5Q users?} Although Figure~\ref{fig:dayedge} demonstrates that internal
edges always outnumber external edges, the reality of the situation is more complicated
when the edges are separated by OSN. 

Figure~\ref{fig:ratioive} plots the ratio of internal to external edges over time for Xiaonei
and 5Q. Initially, users on both OSNs favor creating internal edges ({\em i.e.} the ratio is
$>$1). However, by day 16, the ratio for 5Q users starts to permanently favor external
edges. The reason for this strange result is that Xiaonei users create more than twice as many
edges than 5Q users. In our dataset Xiaonei users create 3.9 million internal edges,
while 5Q users only create 1.5 million. However, unlike internal edges,
external edges affect the statistics for {\em both} groups. Thus, the number of external
edges (2.2 million total in our dataset) is driven by the more active user base. Even though
Xiaonei users create less external edges than internal edges, the number is still proportionally
greater than the number of internal edges created between 5Q users. The ``both'' line in
Figure~\ref{fig:ratioive} is always $>$1 because Xiaonei users create more edges overall, which
weights the average upwards.

Figure~\ref{fig:rationve} plots the ratio of edges to new users versus external edges over
time for Xiaonei and 5Q. This plot reveals that the inflection point where users switch from
preferring external edges to new edges is different for the two OSNs. The ratio becomes $\ge$1
for Xiaonei 5 days after the merge, whereas 5Q takes 32 days. Despite these differences, both
OSNs demonstrate the same overall trend for the ratio to eventually tip heavily in favor of
edges to new users. 

\para{Distance Between Xiaonei and 5Q.} Finally, we examine the practical consequences of
edge creation between Xiaonei and 5Q. Our goal is to answer the question: {\em at what point
do Xiaonei and 5Q become so interconnected that they can no longer be considered separate graphs?}

To answer this question, we calculate the distance, in hops, between users in each
group. Intuitively, the distance between the groups should decrease over time as 1)
more external edges are created, and 2) more internal edges increase the connectivity
of users with external edges. In our experiments, we select 1,000 random users from each
OSN on each day after the merge and calculate the shortest path from each of them to {\em any}
user in the opposite OSN. Thus, the lowest value possible in this experiment is 1, {\em e.g.}
the randomly selected user has an external edge directly to a user in the opposite OSN.
New users and edges to new users are not considered in these tests.

Figure~\ref{fig:graphdistance} shows that the average path length between the two OSNs rapidly
declines over time. Although average path lengths for both OSNs initially start above
3 hops, within 47 days average path lengths are $<$2. Path lengths from Xiaonei
to 5Q are uniformly shorter, and by the end of the experiment the average path length
is $<$1.5.

The distance between Xiaonei and 5Q rapidly approaches an asymptotic lower bound in
Figure~\ref{fig:graphdistance}. Once this bound is reached, it is apparent that the graphs
can become no closer together. Thus, we conclude that by day 50, when both lines
begin to flatten and approach the lower bound, Xiaonei and 5Q can no longer be considered
separate OSNs. These results demonstrate how quickly the two disjoint OSNs can merge into
a single whole, even when edge creation is biased in favor of internal edges
(see Figure~\ref{fig:ratioive}).

\subsection{Summary of Results}

Our analysis of the network merge produces several high-level conclusions:

\begin{packed_itemize}
\item {\em There were a large number of duplicate accounts between Xiaonei and 5Q that become
inactive immediately after the merge.}
\item {\em Edges to new nodes quickly become the driving force behind edge creation.}
\item {\em Despite user's preference against external edges, Xiaonei and 5Q very quickly merge
into a single, well connected graph.}
\end{packed_itemize}

We also observe that the network merge alters user's edge creation patterns for a short time (until
equilibrium is restored):

\begin{packed_itemize}
\item {\em The total number of edges created per day increases, driven by the sudden appearance
of so many new users.}
\item {\em Users individual preferences for internal/external edges changes drastically in the
days following the merge.}
\item {\em Xiaonei users are more active than 5Q users. Thus, the external edges created between
Xiaonei and 5Q force 5Q users to become more active than they normally would be.}
\end{packed_itemize}

%% file: related.tex
\section{Related Work}


\para{Dynamic OSN Measurement.}
Several studies have measured basic dynamic properties of
graphs.  \cite{leskovec2005graphs} analyzed four citation and patent graphs,
and proposed the forest fire model to explain the observed graph densification
and shrinking diameter. \cite{leskovec2008microscopic} studied details of
dynamics in four OSNs to confirm preferential attachment and triangle closure
features. Similar conclusions were reached by
studies on Flickr~\cite{mislove2008growth} and a social network
aggregator~\cite{garg2009evolution}. 
\cite{kang2010radius} measured network temporal radius and found out that
there is a gelling point to distribution.
In addition, \cite{akoglu2008rtm} measured weighted dynamic
graphs, \cite{ahn2007analysis} analyzed the growth of a Korean OSN, and
\cite{viswanath2009evolution} considered temporal user interactions as graph
edges instead of static friendship. Finally, \cite{guo2009analyzing,kumar2005bursty}
analyzed blogspace dynamics.

Some studies focused on analyzing social network dynamics through explicitly
defined groups~\cite{backstrom2006group,zheleva2009co,kairam2012life} or
disconnected components~\cite{kumar2006structure,mcglohon2008weighted,kang2010patterns}.
\cite{kumar2005bursty} tried to identify blog communities and detect bursts
in different temporal snapshots.
\cite{palla2007quantifying} utilized the clique percolation
method~\cite{derenyi2005clique} to identify overlapping community dynamics
in mobile and citation graphs.  Unlike these studies, our work focuses on the
evolution of implicit communities in a densely connected, large-scale social
graph.
 
\para{Dynamic Community Detection and Tracking Algorithms.}
There are two approaches to detecting and tracking dynamic communities. One approach
is to minimize the self-defined temporal
cost of communities between snapshots. \cite{tantipathananandh2007framework}
proved that this problem is NP-hard and then several
works~\cite{tantipathananandh2007framework,tantipathananandh2009constant,lin2008facetnet}
proposed approximation algorithms. However, these algorithms only scale to graphs
with thousands of
nodes. \cite{sun2007graphscope} and \cite{kim2009particle} propose dynamic
community detection algorithms that scale to graphs with hundreds of thousands of
nodes. The drawback of \cite{sun2007graphscope} is that it cannot track individual
community evolution.

The other approach is to match communities detected by static community
detection algorithms across temporal snapshots. \cite{greene2010tracking} maps
communities between snapshots if their similarity is higher than a
threshold. \cite{asur2009event,takaffoli2010framework}
tracks communities between snapshots based on critical community
events. These algorithms do not consider any temporal correlation of
communities when they detect communities between snapshots.

%% file: conclusion.tex
\section{Conclusion}

This work presents a detailed analysis of user dynamics in a large online
social network, using a dataset that covers the creation of 19 million users
and 199 million edges over a 25 month period.  More specifically, we focus on
analyzing edge dynamics at different levels of scale, including dynamics at
the level of individual users, dynamics involving the merge and split of
communities, and dynamics involving the merging of two independent online
social networks.

Our analysis produced a number of interesting findings of dynamics at
different scales.  First, at the individual node level, we found that the
preferential attachment model gradually weakens in impact as the network
grows and matures.  In fact, edge creation in general becomes
increasingly driven by connections between existing nodes as the network
matures, even as node growth keeps pace with the growth in overall network
size.  Second, at the community level, we use an incremental version of the
popular Louvain community detection algorithm to track communities across
snapshots.  We empirically analyze the birth, growth, and death of communities
across merge and split events, and show that community merges can be
predicted with reasonable accuracy using structural features and dynamic
metrics such as acceleration in community size.  Finally, we analyze detailed
dynamics following a unique event merging two comparably-sized social
networks, and observe that its impact, while significant in the short term,
quickly fades with the constant arrival of new nodes to the system.

While our results from Renren may not generalize to all social networks, our
analysis provides a template for understanding the dynamic processes that are
active at different scales in many complex networks. A
significant take-away from our work is that the actions of individual users
are not only driven by dynamic processes at the node-level, but are also
significantly influenced by events at the community and network levels.  A
comprehensive understanding or model of an evolving network must account for
changes at the network and community levels and their impact on individual
users. 

%% file: renren.bbl
\begin{thebibliography}{10}

\bibitem{ahn2007analysis}
{\sc Ahn, Y., Han, S., Kwak, H., Moon, S., and Jeong, H.}
\newblock Analysis of topological characteristics of huge online social
  networking services.
\newblock In {\em Proc of {WWW}\/} (2007).

\bibitem{akoglu2008rtm}
{\sc Akoglu, L., McGlohon, M., and Faloutsos, C.}
\newblock {RTM}: Laws and a recursive generator for weighted time-evolving
  graphs.
\newblock In {\em Proc. of {ICDM}\/} (2008).

\bibitem{asur2009event}
{\sc Asur, S., Parthasarathy, S., and Ucar, D.}
\newblock An event-based framework for characterizing the evolutionary behavior
  of interaction graphs.
\newblock {\em ACM TKDD 3}, 4 (2009), 16.

\bibitem{backstrom2006group}
{\sc Backstrom, L., Huttenlocher, D., Kleinberg, J., and Lan, X.}
\newblock Group formation in large social networks: membership, growth, and
  evolution.
\newblock In {\em Proc. of {KDD}\/} (2006).

\bibitem{barabasi1999emergence}
{\sc Barab{\'a}si, A., and Albert, R.}
\newblock Emergence of scaling in random networks.
\newblock {\em Science 286}, 5439 (1999), 509.

\bibitem{blondel10008fast}
{\sc Blondel, V., Guillaume, J., Lambiotte, R., and Lefebvre, E.}
\newblock Fast unfolding of communities in large networks, 2008.
\newblock {\em J. Stat. Mech\/}.

\bibitem{clauset2004finding}
{\sc Clauset, A., Newman, M., and Moore, C.}
\newblock Finding community structure in very large networks.
\newblock {\em Physical review E 70}, 6 (2004).

\bibitem{derenyi2005clique}
{\sc Der{\'e}nyi, I., Palla, G., and Vicsek, T.}
\newblock Clique percolation in random networks.
\newblock {\em Physical review letters 94\/} (2005).

\bibitem{garg2009evolution}
{\sc Garg, S., Gupta, T., Carlsson, N., and Mahanti, A.}
\newblock Evolution of an online social aggregation network: an empirical
  study.
\newblock In {\em Proc. of {IMC}\/} (2009).

\bibitem{greene2010tracking}
{\sc Greene, D., Doyle, D., and Cunningham, P.}
\newblock Tracking the evolution of communities in dynamic social networks.
\newblock In {\em Proc. of ASONAM\/} (2010).

\bibitem{guo2009analyzing}
{\sc Guo, L., Tan, E., Chen, S., Zhang, X., and Zhao, Y.}
\newblock Analyzing patterns of user content generation in online social
  networks.
\newblock In {\em Proc. of {KDD}\/} (2009).

\bibitem{renren-imc10}
{\sc Jiang, J., Wilson, C., Wang, X., Huang, P., Sha, W., Dai, Y., and Zhao,
  B.~Y.}
\newblock Understanding latent interactions in online social networks.
\newblock In {\em Proc. of {IMC}\/} (2010).

\bibitem{kairam2012life}
{\sc Kairam, S., Wang, D., and Leskovec, J.}
\newblock The life and death of online groups: predicting group growth and
  longevity.
\newblock In {\em WSDM\/} (2012).

\bibitem{kang2010patterns}
{\sc Kang, U., McGlohon, M., Akoglu, L., and Faloutsos, C.}
\newblock Patterns on the connected components of terabyte-scale graphs.
\newblock In {\em Proc. of ICDM\/} (2010).

\bibitem{kang2010radius}
{\sc Kang, U., Tsourakakis, C., and Faloutsos, C.}
\newblock Radius plots for mining tera-byte scale graphs: Algorithms, patterns,
  and observations.
\newblock In {\em Proc. of SDM\/} (2010).

\bibitem{kim2009particle}
{\sc Kim, M., and Han, J.}
\newblock A particle-and-density based evolutionary clustering method for
  dynamic networks.
\newblock {\em Proceedings of the VLDB Endowment 2}, 1 (2009), 622--633.

\bibitem{kumar2005bursty}
{\sc Kumar, R., Novak, J., Raghavan, P., and Tomkins, A.}
\newblock On the bursty evolution of blogspace.
\newblock {\em World Wide Web 8}, 2 (2005), 159--178.

\bibitem{kumar2006structure}
{\sc Kumar, R., Novak, J., and Tomkins, A.}
\newblock Structure and evolution of online social networks.
\newblock In {\em Proc. of {KDD}\/} (2006).

\bibitem{kwak2009mining}
{\sc Kwak, H., Choi, Y., Eom, Y., Jeong, H., and Moon, S.}
\newblock Mining communities in networks: a solution for consistency and its
  evaluation.
\newblock In {\em Proc. of {IMC}\/} (2009).

\bibitem{leskovec2008microscopic}
{\sc Leskovec, J., Backstrom, L., Kumar, R., and Tomkins, A.}
\newblock Microscopic evolution of social networks.
\newblock In {\em Proc. of {KDD}\/} (2008).

\bibitem{leskovec2005graphs}
{\sc Leskovec, J., Kleinberg, J., and Faloutsos, C.}
\newblock Graphs over time: densification laws, shrinking diameters and
  possible explanations.
\newblock In {\em Proc. of {KDD}\/} (2005).

\bibitem{lin2008facetnet}
{\sc Lin, Y., Chi, Y., Zhu, S., Sundaram, H., and Tseng, B.}
\newblock Facetnet: a framework for analyzing communities and their evolutions
  in dynamic networks.
\newblock In {\em Proc of {WWW}\/} (2008).

\bibitem{linkedin}
{\sc {LinkedIn Infrastructure Team}}.
\newblock Data infrastructure at linkedin.
\newblock In {\em Proc. of ICDE\/} (2012).

\bibitem{mcglohon2008weighted}
{\sc McGlohon, M., Akoglu, L., and Faloutsos, C.}
\newblock Weighted graphs and disconnected components: patterns and a
  generator.
\newblock In {\em Proc. of {KDD}\/} (2008).

\bibitem{mislove2008growth}
{\sc Mislove, A., et~al.}
\newblock Growth of the flickr social network.
\newblock In {\em Proc. of {WOSN}\/} (2008).

\bibitem{newman2004finding}
{\sc Newman, M., and Girvan, M.}
\newblock Finding and evaluating community structure in networks.
\newblock {\em Physical review E 69}, 2 (2004).

\bibitem{palla2007quantifying}
{\sc Palla, G., Barabasi, A., and Vicsek, T.}
\newblock Quantifying social group evolution.
\newblock {\em Nature 446}, 7136 (2007), 664--667.

\bibitem{sun2007graphscope}
{\sc Sun, J., Faloutsos, C., Papadimitriou, S., and Yu, P.}
\newblock Graphscope: parameter-free mining of large time-evolving graphs.
\newblock In {\em Proc. of {KDD}\/} (2007).

\bibitem{takaffoli2010framework}
{\sc Takaffoli, M., et~al.}
\newblock A framework for analyzing dynamic social networks.
\newblock {\em Applications of Social network Analysis (ASNA)\/} (2010).

\bibitem{tantipathananandh2009constant}
{\sc Tantipathananandh, C., and Berger-Wolf, T.}
\newblock Constant-factor approximation algorithms for identifying dynamic
  communities.
\newblock In {\em Proc. of {KDD}\/} (2009).

\bibitem{tantipathananandh2007framework}
{\sc Tantipathananandh, C., Berger-Wolf, T., and Kempe, D.}
\newblock A framework for community identification in dynamic social networks.
\newblock In {\em Proc. of {KDD}\/} (2007).

\bibitem{viswanath2009evolution}
{\sc Viswanath, B., Mislove, A., Cha, M., and Gummadi, K.}
\newblock On the evolution of user interaction in facebook.
\newblock In {\em Proc. of {WOSN}\/} (2009).

\bibitem{Wakita2007}
{\sc Wakita, K., and Tsurumi, T.}
\newblock Finding community structure in mega-scale social networks.
\newblock {\em CoRR abs/cs/0702048\/} (2007).

\bibitem{interaction}
{\sc Wilson, C., Boe, B., Sala, A., Puttaswamy, K. P.~N., and Zhao, B.~Y.}
\newblock User interactions in social networks and their implications.
\newblock In {\em Proc. of EuroSys\/} (April 2009).

\bibitem{zachary1977information}
{\sc Zachary, W.}
\newblock An information flow model for conflict and fission in small groups.
\newblock {\em Journal of anthropological research\/} (1977), 452--473.

\bibitem{svm2001}
{\sc Zhang, T.}
\newblock An introduction to support vector machines and other kernel-based
  learning methods.
\newblock {\em AI Magazine 22}, 2 (2001), 103.

\bibitem{zheleva2009co}
{\sc Zheleva, E., Sharara, H., and Getoor, L.}
\newblock Co-evolution of social and affiliation networks.
\newblock In {\em Proc. of {KDD}\/} (2009).

\end{thebibliography}
